\documentclass[preprint,aps,tightenlines,showkeys,nofootinbib,superscriptaddress]{revtex4}
\usepackage{latexsym}
\usepackage[dvips]{graphicx}
\usepackage[dvips]{color}
\newcommand{\beq}{\begin{eqnarray}}
\newcommand{\eeq}{\end{eqnarray}}
\newcommand{\bea}{\begin{eqnarray*}}
\newcommand{\eea}{\end{eqnarray*}}
\newcommand{\eq}{eqnarray}

\newcommand{\al}{{\alpha}}

\newcommand{\ci}{\cite}
\newcommand{\ga}{{\gamma}}
\newcommand{\Ga}{{\Gamma}}
\newcommand{\ep}{{\epsilon}}

\newcommand{\de}{{\delta}}
\newcommand{\De}{\Delta}

\newcommand{\la}{{\lambda}}
\newcommand{\La}{{\Lambda}}

\newcommand{\si}{{\sigma}}
\newcommand{\Si}{{\Sigma}}
\newcommand{\ka}{{\kappa}}
\newcommand{\om}{{\omega}}
\newcommand{\Om}{{\Omega}}
\newcommand{\pa}{{\partial}}
\newcommand{\no}{{\nonumber}}
\newcommand{\f}{\frac}

\newcommand{\ra}{\rightarrow}
\newcommand{\lra}{\leftrightarrow}

\newcommand{\sqg}{\sqrt{g}}

\begin{document}

\preprint{arXiv:{2001.02556v4} [hep-th]}

\title{The Hamiltonian Dynamics of
Ho\v{r}ava Gravity }

\author{Deniz O. Devecio\u{g}lu \footnote{E-mail address: dodeve@gmail.com}}
\affiliation{School of Physics, Huazhong University of Science and Technology,
Wuhan, Hubei,  430074, China }

\author{Mu-In Park \footnote{E-mail address: muinpark@gmail.com, Corresponding author} \footnote{Former address: Research Institute for Basic Science, Sogang University,
Seoul, 121-742, Korea}}
\affiliation{ Center for Quantum Spacetime, Sogang University,
Seoul, 121-742, Korea }\date{\today}

\begin{abstract}
We consider the Hamiltonian formulation of Ho\v{r}ava gravity in {\it arbitrary} dimensions, which has been proposed as a renormalizable gravity model for
quantum gravity without the ghost problem. We study the {\it full} constraint analysis of the {\it non-projectable} Ho\v{r}ava gravity whose potential, ${\cal V}(R)$, is an arbitrary function of the (intrinsic) Ricci scalar $R$ but without the extension terms which depend on the proper
acceleration $a_i$.
We find that there exist generally three distinct cases of this theory, {\bf A}, {\bf B}, and {\bf C}, depending on (i) whether the
Hamiltonian constraint generates new (second-class) constraints
or just fixes the associated Lagrange multipliers,
or (ii) whether the IR Lorentz-deformation parameter $\la$ is at the conformal point
or not.
It is found that,
for
Cases {\bf A} and {\bf C},
the dynamical degrees of freedom are the same as in general relativity,
while, for Case {\bf B}, there is {\it one additional phase-space degree of freedom}, representing
an extra (odd) scalar graviton mode. 
This would achieve the dynamical consistency
of a restricted model at the {\it fully non-linear} level and be a positive result in resolving the long-standing debates about the extra graviton modes of the Ho\v{r}ava gravity. Several exact solutions are
also studied
as some explicit examples of the new constraints. The structure of
the newly obtained, ``extended" constraint algebra seems to be generic to Ho\v{r}ava gravity
and its general proof would be a challenging problem. Some other challenging problems,
which include
the path integral quantization and the Dirac bracket quantization
are discussed also.
\end{abstract}



\maketitle

\newpage

\section{Introduction}

11 years ago, Ho\v{r}ava proposed a renormalizable, higher-derivative gravity theory,
without the ghost problem in the usual {\it tensor} graviton 
modes,
which reduces to Einstein gravity in low energy (IR) at the full action level but
with improved high-energy (UV) behaviors, by abandoning the Lorentz symmetry from
non-equal-footing treatment of space and time
\ci{Hora:0812,Hora}. However, due to absence of the full diffeomorphisms ({\it Diff}), extra graviton modes
can be expected generally and there have been questions
about the recovery of general relativity (GR) in IR, and more generally, the consistency of Ho\v{r}ava gravity
\ci{Char,Li,Blas:0906,Koba,Blas:0909,Park:0910a,Koya,Park:0910b,Henn,Gong,Pons:2010,
Bell:1004,Bell:1010,Loll:2014}.

In this paper, we reconsider the Hamiltonian formulation of the {\it non-projectable}
Ho\v{r}ava gravity whose potential, ${\cal V}(R)$, is an arbitrary function of the (intrinsic)
Ricci scalar $R$ in arbitrary dimensions but without the extension terms which depend on the proper acceleration $a_i=\pa_i N/N$, for simplicity. We study the {\it full} constraint analysis and find that
there exist generally three distinct cases of this theory, {\bf A}, {\bf B}, and {\bf C}, depending on
(i) whether the Hamiltonian constraint generates new
(second-class) constraints,
 or just fixes the associated Lagrange multipliers,
according to Dirac's method \ci{Dira}, or
(ii) whether the IR Lorentz-deformation
parameter $\la$ is at the conformal point, $\la=1/D$,
 or not.
It is found that,
for Cases {\bf A} and {\bf C}, the dynamical degrees of freedom are
the same as in general relativity, while, for Case {\bf B}, there is {\it one additional phase-space
degree of freedom}, representing an extra (odd) scalar graviton mode.
This would achieve the dynamical consistency
of a restricted model at the {\it fully non-linear} level and be a positive result in resolving the long-standing debates about the extra
graviton modes of the Ho\v{r}ava gravity.

The organization of the paper is as follows. In Sec. II, we consider the set-up for the
Hamiltonian formulation of the non-projectable Ho\v{r}ava gravity in arbitrary dimensions.
In Sec. III, we consider Dirac's constraint analysis when $\la \neq 1/D$ for the IR Lorentz-deformation parameter $\la$ and study two cases, {\bf A} and {\bf B}, depending
on whether the Hamiltonian constraint generates new (second-class)
constraints
or just fixes the associated Lagrange multipliers.
In Sec. IV, we consider the third case, {\bf C},
when $\la=1/D$, which is at a
conformal point
and generates new (second-class) constraints similar to Case {\bf A}.
We show that the dynamical degrees of freedom are the same as in GR for
Cases {\bf A} and {\bf C}, while there is one additional phase-space degree of freedom for
Case {\bf B} which representing an extra (odd) scalar graviton mode.
In Sec. V, we study
several exact solutions as some explicit examples of the new constraints. In Sec. VI, we
conclude with remarks on several challenging theoretical problems, which include
the path integral quantization and the Dirac bracket quantization.

\section{The Hamiltonian Formulation in Arbitrary Dimensions: Set-up                                                                                                                                                                                                                                                                                                                                                                                                                                                                                                                                                                                                                                                                                                                                                                                                                                                                                                                                                                                                                                                                                                                                                                                                                                                                                                                                                                                                                                                                                                                                                                                                                                                                                                                                                                                                                                                                                                                                                                                                                                                                                                                                                                                                                                                                                                                                                                                                                                                                                                                                                                                                                                                                                                                                                                                                                                                                                                                                                                                                                                                                                                                                                                                                                                                                                                                                                                                                                                                                                                                                                                                                                                                                                                                                                                                                                                                                                                                                                                                                                                                                                                                                                                                                                                                                                                                                                                                                                                                                                                                                                                                                                                                                                                                                                                                                                                                                                                                                                                                                                                                                                                                                                                                                                                                                                                                                                                                                                                                                                                                                                                                                                                                                                                                                                                                                                                                                                                                                                                                                                                                                                                                                                                                                                                                                                                                                                                                                                                                                                                                                                                                                                                                                                                                                                                                                                                                                                                                                                                                                                                                                                                                                                                                                                                                                                                                                                                                                                                                                                                                                                                                                                                                                                                                                                                                                                                                                                                                                                                                                                                                                                                                                                                                                                                                                                                                                                                                                                                                                                                                                                                                                                                                                                                                                                                                                                                                                                                                                                                                                                                                                                                                                                                                                                                                                                                                                                                                                                                                                                                                                                                                                                                                                                                                                                                                                                                                                                                                                                                                                                                                                                                                                                                                                                                                                                                                                                                                                                                                                                                                                                                                                                                                                                                                                                                                                                                                                                                                                                                                                                                                                                                                                                                                                                                                                                                                                                                                                                                                                                                                                                                                                                                                                                                                                                                                                                                                                                                                                                                                                                                                                                                                                                                                                                                                                                                                                                                                                                                                                                                                                                                                                                                                                                                                                                                                                                                                                                                                                                                                                                                                                                                                                                                                                                                                                                                                                                                                                                                                                                                                                                                                                                                                                                                                                                                                                                                                                                                                                                                                                                                                                                                                                                                                                                                                                                                                                                                                                                                                                                                                                                                                                                                                                                                                                                                                                                                                                                                                                                                                                                                                                                                                                                                                                                                                                                                                                                                                                                                                                                                                                                                                                                                                                                                                                                                                                                                                                                                                                                                                                                                                                                                                                                                                                                                                                                                                                                                                                                                                                                                                                                                                                                                                                                                                                                                                                                                                                                                                                                                                                                                                                                                                                                                                                                                                                                                                                                                                                                                                                                                                                                                                                                                                                                                                                                                                                                                                                                                                                                                                                                                                                                                                                                                                                                                                                                                                                                                                                                                                                                                                                                                                                                                                                                                                                                                                                                                                                                                                                                                                                                                                                                                                                                                                                                                                                                                                                                                                                                                                                                                                                                                                                                                                                                                                                                                                                                                                                                                                                                                                                                                                                                                                                                                                                                                                                                                                                                                                                                                                                                                                                                                                                                                                                                                                                                                                                                                                                                                                                                                                                                                                                                                                                                                                                                                                                                                                   }

In this section, we consider the Hamiltonian formulation of the non-projectable Ho\v{r}ava gravity in arbitrary dimensions. To this ends, we start by considering the ADM decomposition of the metric \ci{Arno}
\begin{\eq}
ds^2=-N^2 dt^2+g_{ij}\left(dx^i+N^i dt\right)\left(dx^j+N^j dt\right)
\end{\eq}
with the arbitrary space-time dependent lapse and shift functions \footnote{For the projectable case, where the lapse function depends only on time coordinate, {\it i.e.}, $N(t)$, there is {\it no} ``local" Hamiltonian constraint so that there is {\it no} {\it smooth} way to recover the usual constraint algebra in GR \ci{Hora:0812,Koba}. Later, we will comment about other alternative formulation which can provide a smooth limit to the non-projectable case.}, $N, N^i$, and induced metric $g_{ij}~ (i,j=1,2, \cdots D)$ on a time-slicing hypersurface $\Sigma_t$. Then, the action on a $(D+1)$-dimensional manifold ${\cal M}$ with the boundary $\partial {\cal M}$ is given by
\begin{\eq}
\label{HL action}
S &=& \int_{\cal M} d t d^{D}{x} \sqrt{g} N \left\{ \frac{2}{\kappa^2}
\left(K_{ij}K^{ij} - \lambda K^2 \right) - {\cal V}[g_{ij}] \right\}+S_{\partial {\cal M}}
\end{\eq}
with an appropriate boundary action \footnote{The explicit form of boundary terms is not essential in this paper and will not be considered in detail. But, due to the similarity to GR or its Lorentz invariant higher-curvature gravity theories, the required boundary terms are also quite close \ci{Deve:2020}, whose IR limit agrees with those of \ci{Donn:2011,Bell:1106}, for example.} $S_{\partial {\cal M}}$ and the potential ${\cal V}[g_{ij}]$, which depends only on the metric $g_{ij}$ and its spatial derivatives \ci{Hora:0812,Hora} \footnote{The {\it UV} Lorentz violation due to higher-spatial-derivative terms in the potential part is originated from the works of E. M. Lifshitz \ci{Lifs} in the study of, so called, ``Lifshitz field theories". On the other hand, the {\it IR} Lorentz violation due to the deformation parameter $\la \neq 1$ in the kinetic part was first studied by B. S. DeWitt in the study of ``canonical quantum gravity" \ci{DeWi}. So, it would be desirable to call the bulk part of action (\ref{HL action}) as {\it DeWitt-Ho\v{r}ava-Lifshitz} ({\it DHL})'s action, more precisely.}. Here,
\begin{\eq}
 K_{ij}=\frac{1}{2N}
\left(\dot{g}_{ij}-\nabla_i N_j-\nabla_jN_i\right)
\end{\eq}
is the extrinsic curvature (the overdot $(~\dot{}~)$ denotes the time derivative) of the hypersurface $\Sigma_t$ and $K \equiv g_{ij} K^{ij}$ denotes its trace.

It is well known that GR is not renormalizable in the conventional way. Its Lorentz invariant higher-curvature modifications have some improvements of UV behaviors but there are more dynamical degrees of freedom than those of GR generally, and the existence of {\it negative} kinetic energy modes in the additional degrees of freedom, called {\it ghost} modes, is unavoidable due to higher-time derivatives \ci{Stel,Buch:1987,Buch:1991}. In order to avoid the {\it possible} problems of ghost degrees of freedom, we do not simply consider the higher-time-derivative terms, like $K^4,(K_{ij}K^{ij})^2$, {\it etc.}, but only consider the second-order time-derivative terms, like $K_{ij}K^{ij}$ and $K^2$, in the kinetic part of the action (\ref{HL action}). Whereas, in order to achieve the renormalizable theory with the improved UV behaviors, we consider the higher-spatial-derivative terms, like $R^2,  R_{ij}R^{ij}$, {\it etc.}, in the potential part
${\cal V}[g_{ij}]$ with the intrinsic Ricci curvature $R_{ij}$ and its trace $R\equiv g_{ij}R^{ij}$. In order that the theory be {\it power-counting} renormalizable, the potential part needs to contain ``$2 \times D$" (spatial) derivatives at least, which is sometimes represented by the dynamical critical exponent, $z=D$ \ci{Hora}. In order that this construction of a renormalizable action is {\it not} spoiled by the mixing of space and time (derivatives) in the general coordinate transformations, we need to further constrain the allowed coordinate transformations into the foliation-preserving diffeomorphisms $({\it Diff}_{\cal F})$,
\begin{\eq}
\label{Diff}
\delta x^i &=&-\zeta^i (t, {\bf x}), ~\delta t=-f(t), \no \\
 \delta
g_{ij}&=&\pa_i\zeta^k g_{jk}+\pa_j \zeta^k g_{ik}+\zeta^k
\pa_k g_{ij}+f \dot g_{ij},\nonumber\\
\delta N_i &=& \pa_i \zeta^j N_j+\zeta^j \pa_j N_i+\dot\zeta^j
g_{ij}+f \dot N_i+\dot f N_i,\no \\
\delta N&=& \zeta^j \pa_j N+f \dot N+\dot f N.
\end{\eq}
In ${\it Diff}_{\cal F}$, each term in the kinetic part is invariant separately and $\la$ can be arbitrary \footnote{For the case $\la = 1/D$, where the theory becomes singular, a separate consideration is needed \ci{Park:0910b,Bell:1106}. We will consider this case later, in Sec. IV.} \ci{DeWi}. If we consider $\la=1$ and ${\cal V}[g_{ij}] =-(2/\ka^2) R-\La$ as in GR, then there is an ``{\it accidental}" symmetry enhancement which mixes each term in the action so that the full {\it Diff} is recovered \ci{Park:0910a}. So, there are two sources of the Lorentz violations, either from the higher-spatial-derivative (UV) terms in the potential part
or from the deformation of kinetic part with an arbitrary $\la$ in IR, generally.

For the potential part, one may consider any function having $2 D$ spatial derivatives for
the power-counting renormalizability, but in this paper we consider only the function of
Ricci curvature scalar, $R$, {\it i.e.,}
${\cal V}[g_{ij}] \equiv {\cal V}(R)$, for simplicity \footnote{The terms of $\nabla^2 R, R_{ij} R^{ij}, etc.$ could also produce some other peculiar
UV behaviors
due to ingenious combinations of terms depending on space-time dimensions,
but we will not consider this possibility in this paper.}.
Then, the first-order formulation of the action (\ref{HL action}) is given by
\begin{\eq}
\label{HL action_1st}
S &=& \int_{{\cal M}} d t d^{D}x  \left\{ \pi^{ij} \dot{g}_{ij}-N {\cal H}-N_i {\cal H}^i-\pa_i \ga^i\right\}
\end{\eq}
with appropriate boundary terms, $\pa_i \ga^i$, the conjugate momenta,
\begin{\eq}
\pi^{ij} \equiv \f{\de S}{\de \dot{g}_{ij}}= \frac{2 \sqrt{g}}{\kappa^2}
\left(K^{ij} - \lambda K g^{ij} \right), \label{mom_def}
\end{\eq}
and
\begin{\eq}
{\cal H}&\equiv&\frac{\kappa^2}{2 \sqrt{g}} \left[\pi^{ij}\pi_{ij} - \left(\f{\lambda}{D \la-1}\right) \pi^2  \right] +\sqrt{g} {\cal V},
\label{H_t}
\\
{\cal H}^i &\equiv&-2 \nabla_j \pi^{ij},
\label{H_i}
\end{\eq}
where
$\pi \equiv g_{ij} \pi^{ij}$. Here, we first consider the case $\la \neq 1/D$  so that ${\cal H}$ in (\ref{H_t}) and the first-order action (\ref{HL action_1st}) are not singular.

The Poisson brackets for the canonical variables are given by
\begin{\eq}
\{ g_{ij} (x), \pi^{kl} (y) \}=\de_{ij}^{kl} \de^D (x-y)
\end{\eq}
with $\de_{ij}^{kl}\equiv (\de_i^k \de_j^l+\de_i^l \de_j^k)/2$.

\section{Constraint Analysis {\it \`{a} la} Dirac ($\la \neq 1/D$) }

The primary constraints of the action (\ref{HL action}) are given by
\begin{\eq}
\Phi^t \equiv \pi_N \approx 0,~\Phi^i \equiv \pi^i \approx 0,
\label{primary_cons}
\end{\eq}
from the definition of conjugate momenta, $\pi_N \equiv \de S/\de \dot{N}$ and $\pi^i \equiv \de S/\de \dot{N_i}$. Here, the weak equality `$\approx$' means that the constraint equations are used only after calculating the Poisson brackets.

The preservation of the primary constraints, $\Phi^\mu \equiv (\Phi^t , \Phi^i)$, {\it i.e.}, $\dot{\Phi}^{\mu} =\{\Phi^{\mu}, H_C \}\approx 0$, as being required by the consistency of the constraints, with the {\it canonical} Hamiltonian,
\begin{\eq}
\label{Hc}
 H_C&=& \int_{\Si_t} d^{D}x  \left\{ N {\cal H}+N_i {\cal H}^i\right\}+H_{\cal B}
\end{\eq}
produces the secondary constraints,
\begin{\eq}
\label{secondary_const}
{\cal H} \approx 0,~~
{\cal H}^i \approx 0.
\end{\eq}
Here, $H_{\cal B}$ is an appropriate boundary Hamiltonian,
$H_{\cal B}=\oint_{\cal B}  d^{D}x~ \hat{n}_i \ga^i$, for the intersection
${\cal B}$ of an arbitrary time-like boundary ($ \hat{n}_i$ is its unit normal) with a time slice $\Si_t$ so that the total
boundaries are $\pa {\cal M}= \Si_{t_f} \cup \Si_{t_i} \cup {\cal B}$.
On the other hand, the dynamical equations for $g_{ij}$ and $\pi^{kl}$ are obtained as follows, neglecting boundary contributions (see Appendix {\bf A} for the details),
\begin{\eq}
\dot{g}_{ij}&=&\{ g_{ij}, H_C \} = \f{\de H_C}{\de \pi^{ij}} \no \\
&=&\left( \frac{\kappa^2}{2} \right) \f{2 N}{ \sqrt{g}} \left(\pi_{ij}-\widetilde{\la}  g_{ij} \pi \right) + \nabla_i N_j + \nabla_j N_i, \label{g_dot}\\
\dot{\pi}^{ij}&=&\{ \pi^{ij}, H_C \}= -\f{\de H_C}{\de g_{ij}} \no \\
&=&\left( \frac{\kappa^2}{2} \right) \f{ N}{ \sqrt{g}} \left[\f{1}{2} g^{ij} \left( \pi_{mn} \pi^{mn}-\widetilde{\la} \pi^2 \right)
-2 \left( \pi^{im} \pi^j_m-\widetilde{\la} \pi \pi^{ij} \right) \right] \no \\
 &&-N \sqrt{g} \left[  \f{ 1}{2} g^{ij} {\cal V}(R)-R^{ij} {\cal V}'(R) \right]
 -\sqrt{g} \left[ \nabla^i \nabla^j \left( N {\cal V}'(R) \right)-g^{ij} \nabla_m \nabla^m \left( N {\cal V}'(R)\right) \right] \no \\
&& + \nabla_m \left(N^m \pi^{ij} \right) -\left(\nabla_m N^i\right) \pi^{jm}-\left(\nabla_m N^j\right) \pi^{im}, \label{pi_dot}
\end{\eq}
where $\widetilde{\la}\equiv\la/(D \la-1)$ and $(~)' \equiv d( ~)/dR$.

With the primary constraints in (\ref{primary_cons}), one can consider the {\it extended} Hamiltonian with the Lagrange multipliers $u_{\mu}$,
\begin{\eq}
\label{HE}
 H_E&=&H_C + \int_{\Si_t} d^{D}x~(   u_{\mu} \Phi^{\mu} ),
\end{\eq}
from the arbitrariness in the equations of motion, due to the primary constraints.
Then, after tedious computations, we obtain the following constraint algebra (see Appendix {\bf A} for the details),
\begin{\eq}
 \{ {\cal H} (x),{\cal H} (y) \}&=&C^i (x) \nabla^x_i \de^D (x-y)-C^i (y) \nabla^y_i \de^D (x-y), \label{HtHt}\\
 \{ {\cal H} (x),{\cal H}_i (y) \}&=&-{\cal H} (y) \nabla^y_i
 \de^D (x-y), \label{HtHi}\\
 \{ {\cal H}_i (x),{\cal H}_j (y) \}&=&{\cal H}_i  (y) \nabla^x_j \de^D (x-y)+{\cal H}_j  (x) \nabla^x_i \de^D (x-y), \label{HiHj}
\end{\eq}
or, for the smeared constraints, $\left< \eta  {\cal H} \right> \equiv \int d^D x~ \eta {\cal H}$, {\it etc.}, with the smearing functions $\eta$ and $\eta_i$,
\begin{\eq}
 \{ \left<\eta {\cal H} \right>,\left<\zeta {\cal H}\right> \}&=& \left<  \left( \eta \nabla_i \zeta-\zeta \nabla_i \eta \right) C^i \right> ,
 \label{HtHt_sm}\\
 \{ \left< \eta {\cal H} \right>,\left< \zeta^i {\cal H}_i\right>\}
 &=&- \left<  \zeta^i \nabla_i \eta {\cal H} \right>  ,  \label{HtHi_sm} \\
 \{ \left< \eta^i {\cal H}_i \right>,\left< \zeta^j {\cal H}_j \right> \}&=& \left< \left( \eta^i \nabla_i \zeta^j-\zeta^i \nabla_i \eta^j \right) {\cal H}_j \right>  , \label{HiHj_sm}
\end{\eq}
where \footnote{From the Jacobi's identity, one may obtain some
{non-trivial} relations about $C^i$. For example, from $\{ \{ \left<\eta {\cal H} \right>,\left<\zeta {\cal H}\right> \}, \left<\rho {\cal H}\right> \}+\{ \{ \left<\zeta {\cal H} \right>,\left<\rho {\cal H}\right> \}, \left<\eta {\cal H}\right> \}+\{ \{ \left<\rho {\cal H} \right>,\left<\eta {\cal H}\right> \}, \left<\zeta {\cal H}\right> \}=0$, one obtains ``$\left<  \left( \eta \nabla_i \zeta-\zeta \nabla_i \eta \right) \Theta^i|_{N \ra \rho} \right>$ +(\mbox{cyclic permutations about} $\eta, \zeta, \rho)=0$", where $\Theta^i$ is defined by $\Theta^i \equiv \{C^i, \left<N {\cal H}\right> \}$ and its explicit form is given by (\ref{Theta_i}), wherein the momentum constraint, ${\cal H}_i \approx 0$, is imposed. }
\begin{\eq}
 C^i & \equiv &  \sqg   {{\cal V}' }^2 (R)  \nabla_j \left(\f{K^{ij}}{ {\cal V}' (R) }
\right) 
\no \\
&=&\left(\f{\ka^2}{2} \right) 2  {{\cal V}' }^2 (R) \nabla_j \left(\f{\pi^{ij}-
\widehat{\la} g^{ij} \pi}{ {\cal V}' (R) } \right) 
\no \\
&=&\left(-\f{\ka^2}{2} \right) \left[\left({\cal H}^i + 2 \widehat{\la} \nabla^i \pi \right) {\cal V}' (R) +2 \left(\pi^{ij}-\widehat{\la} g^{ij} \pi \right) \nabla_j  {\cal V}' (R)  \right]
\label{Ci_1}
\end{\eq}
with $\widehat{\la}\equiv (\la-1)/(D \la -1)$.
If we consider $- {\cal V}(R) \equiv \La+\xi R+ \al R^n$ as a typical example with an
arbitrary power $n$,
we can obtain
\begin{\eq}
C^i=\left(\f{\ka^2}{2} \right) \left[\left({\cal H}^i + 2 \widehat{\la} \nabla^i \pi \right)\left(\xi+\al n R^{n-1} \right) +2 \left(\pi^{ij}-\widehat{\la} g^{ij} \pi \right) \al n \nabla_j R^{n-1}   \right].
\label{Ci_2}
\end{\eq}
Note that, when the higher-derivative contributions are absent, {\it i.e.,} $\al=0$, (\ref{Ci_2}) reduces to $C^i =({\ka^2}/{2}) \xi {\cal H}^i \approx 0$ and the Hamiltonian constraint, ${\cal H} \approx 0$, becomes the first-class constraint as in GR, for either (i) $\la=1$, {\it i.e.}, $\widehat{\la}=0$, or (ii) $\pi \approx 0$, representing the {\it maximal} slicing, for an arbitrary $\la$ \ci{Arno}. However, when the higher-derivative terms are present, $C^i$ does not vanish {\it generally}, so that `` {\it the Hamiltonian constraint, ${\cal H} \approx 0$, becomes the second-class constraint} " even for the maximal slicing, $\pi\approx 0$,
due to the genuine dynamical degrees of freedom in $\pi^{ij}$,
other than the trace part, $\pi$.

On the other hand, we note that the momentum constraint, ${\cal H}_i\approx 0$, in (\ref{secondary_const}) has the same {\it canonical} form as in GR with no higher-derivative corrections \footnote{This is a key observation for the proof of Birkhoff's theorem in Ho\v{r}ava gravity \cite{Deve:2019}, in contrast to other general higher-curvature gravities \ci{Buch:1987,Buch:1991}. But the IR Lorentz-deformation parameter $\la$ enters still in the momentum constraint through the relation (\ref{mom_def}). } so that we also have the same constraint algebra for ${\cal H}_i$, (\ref{HiHj}) or (\ref{HiHj_sm}), which representing the spatial {\it Diff}  generator,
\begin{\eq}
\de_{\zeta} g_{ij}&=&\{ g_{ij}, \left< \zeta^k {\cal H}_k \right> \}
=\f{\de \left< \zeta^k {\cal H}_k\right>}{\de \pi^{ij}}\no \\
&=&\nabla_i \zeta_j +\nabla_j \zeta_j, \\
\de_{\zeta} \pi^{ij}&=&\{ \pi^{ij}, \left< \zeta^k {\cal H}_k \right> \}=-\f{\de \left< \zeta^k {\cal H}_k\right>}{\de g_{ij}}\no \\
&=& \nabla_m \left(\zeta^m \pi^{ij} \right) -\left(\nabla_m \zeta^i\right) \pi^{jm}-\left(\nabla_m \zeta^j\right) \pi^{im},
\end{\eq}
as in GR.
Moreover, the algebras (\ref{HtHi}) and (\ref{HiHj}) show that
\begin{\eq}
\de_{\zeta}{\cal H}&=&\{ {\cal H}, \left< \zeta^k {\cal H}_k \right> \}= \nabla_k \left(\zeta^k {\cal H} \right), \\
\de_{\zeta} {\cal H}_i &=&\{ {\cal H}_i, \left< \zeta^k {\cal H}_k \right> \}=\nabla_k \left(\zeta^k {\cal H}_i \right)+\left(\nabla_i \zeta^k\right) {\cal H}_k,
\end{\eq}
which tells that ${\cal H}$ and ${\cal H}_i$ behave as, under the spatial {\it Diff}, the scalar and vector densities, respectively, as in GR.

Using the above constraint algebra, one can easily find that the preservation of the secondary constraints give
\begin{\eq}
\dot{{\cal H}}&=&\{{\cal H} , H_E \} \no \\
&=&\f{1}{N} \nabla_i (N^2 C^i) +\nabla_i (N^i {\cal H})
\approx \f{1}{N} \nabla_i (N^2 \widetilde{C}^i) , \label{H_t_dot} \\
\dot{{\cal H}_i}&=&\{{\cal H}_i , H_E \} \no \\
&=&{\cal H} \nabla_i N +\nabla_j (N^j {\cal H}_i)+{\cal H}_j \nabla_i N^j \approx 0,
\label{H_i_dot}
\end{\eq}
which produces the {\it tertiary} constraint,
\begin{\eq}
\widetilde{\Omega} \equiv \nabla_i (N^2 \widetilde{C}^i)
\approx 0,
\label{Omega_const}
\end{\eq}
where
\begin{\eq}
\widetilde{C}^i \equiv C^i |_{{\cal H}_i \approx 0}  =\left(\f{\ka^2}{2} \right) \left[ 2 \widehat{\la} \nabla^i \pi \left(\xi+\al n R^{n-1} \right) +2 \left(\pi^{ij}-\widehat{\la} g^{ij} \pi \right) \al n \nabla_j R^{n-1}   \right]
\label{Ci_3}
\end{\eq}
from the preservation of ${\cal H} \approx 0$ in (\ref{H_t_dot}), excluding the trivial case of
$N=0$ for all space-time. 
Here we note that, in the above computations, there are no contributions from
the multiplier terms in $H_E$ so that we need to consider further steps of preserving the
constraints until the multipliers are determined.

Now, one more step of preserving the new tertiary constraint, $\widetilde{\Omega} \approx 0$, gives
\begin{\eq}
\dot{\widetilde{\Omega}}&=& \{\widetilde{\Omega}, H_E\} \no \\
&=&\{\widetilde{\Omega}, H_C\}+2 \widetilde{\Omega} \left( \f{u_t}{N} \right)
+2 \widetilde{C}^i N^2 \nabla_i \left( \f{u_t}{N} \right)
 \no \\
&\approx & \{\widetilde{\Omega}, H_C\}+2 \widetilde{C}^i N^2 \nabla_i \left( \f{u_t}{N} \right) \approx 0.
\label{Om_dot}
\end{\eq}
Then, there are two different constraint systems, with different subsequent procedures,
depending on whether
$\widetilde{C}^i = 0$ or $\widetilde{C}^i \neq 0$.\\

{\bf A}. Case $\widetilde{C}^i = 0$:
In this case \footnote{Here, the condition $\widetilde{C}^i = 0$ does not necessarily mean the
constraint equation, which is stronger than the original constraint
(\ref{Omega_const}). Actually, the constraint (\ref{Omega_const}) implies that
all the components of $\widetilde{C}^i$ are not independent so that the condition
$\widetilde{C}^i = 0$ may be subject to more fundamental conditions or constraints.}, (\ref{Om_dot})
does not determine the multiplier $u_t$ but reduces to
\begin{\eq}
\dot{\widetilde{\Omega}} &\approx& \{\widetilde{\Om} , H_C \} \no \\
 &\approx& \nabla_i \left( N^2 \widetilde{\Theta}^i \right)
 \equiv \widetilde{\Xi} \approx 0
 \label{Phi_constraint}
\end{\eq}
for preserving the tertiary constraint $\widetilde{\Om}$ with
\begin{\eq}
\widetilde{\Theta}^i &\equiv& \{\widetilde{C}^i, \left<N {\cal H} \right> \} \no \\
&=&-\left(\f{\ka^2}{2} \right)^2 \f{2 \widehat{\la}}{(\la D-1) \sqrt{g}}
{\cal V}'(R)
\left\{
\left[ (2 \la+1) g^{ij} \pi-2 (\la D-1) \pi^{ij}\right]
N \nabla_j \pi+\pi^2 \nabla^i N
\right\} \no \\
&&- \left(\f{\ka^2}{2} \right) 2 \widehat{\la} (D-1)  \sqrt{g} {\cal V}'(R) \nabla^i
\left[ \left(\xi R+\f{ \La D}{D-1}
 + \frac{D-n}{D-1} \alpha R^{n} \right) N +\nabla^2 (N\mathcal{V}^{\prime})  \right] \no \\
&&- \left(\f{\ka^2}{2} \right) \left[ 2 \widehat{\la} \nabla_i \pi \{{\cal V}'(R), \left<N {\cal H} \right> \}+
\{2 (\pi^{ij}-\widehat{\la} g^{ij} \pi ) \nabla_j {\cal V}'(R),\left<N {\cal H} \right> \}\right],
\label{Theta_i}
\end{\eq}
where $-{\cal V}'(R) \equiv \left(\xi +\al n R^{n-1} \right)$.
Here, the higher-derivative contributions come, either from the constraint
(\ref{Omega_const}), $\widetilde{\Om}$, {\it i.e.}, $\widetilde{C}^i$ in (\ref{Ci_3}), or from the Hamiltonian constraint ${\cal H}$ in (\ref{H_t}) and (\ref{Hc}). One can compute the explicit forms of the higher-derivative contributions with the help of (\ref{g_dot}) and (\ref{pi_dot}) (see Appendix {\bf B} for some more details; see also \ci{Bell:1010} for the case of $n=2$) but, due to its messy expression, we will not consider the explicit forms in the analysis below unless it is crucial.

Then, one more time-evolution of the new constraint,
$\widetilde{\Xi}\approx 0$, will read,
\begin{\eq}
\dot{\widetilde{\Xi}}&=&\{\widetilde{\Xi}, H_E \} \no \\
&=&\{\widetilde{\Xi}, H_C \}+\nabla_i \left(N^2 ~ \widetilde{\Theta}^i|_{N \ra u_t}+2N u_t \widetilde{\Theta}^i   \right) \approx 0.
\label{Phi_time_evol_eq}
\end{\eq}
After a long computation, we obtain (see also Appendix {\bf B} for the details) \footnote{{For convenience, we consider the modified momentum constraint \ci{Klus:2010}, $\overline{\cal H}^i \equiv {\cal H}^i+\pi_N \nabla^i N$, by redefining the multiplier, $u_t \ra u_t-N_i \nabla^i N$ in (\ref{HE}) so that $H_E=\overline{H}_C+ \left< (u_t-N_i \nabla^i N) \pi_N+u_i \pi^i  \right>$, $\overline{H}_C \equiv \left<  N {\cal H}+N_i \overline{\cal H}^i \right> $ and $\{ \Xi, \left< N_j {\cal H}^j\right>\}=\nabla_j ( N^j \widetilde{\Xi} )$. In this way, one can compactly collect all $N_i$-dependent terms in the left hand side, up to the weakly vanishing term, $\nabla_j ( N^j \widetilde{\Xi} )\approx 0$. Interestingly, this modified constraint $\overline{\cal H}^i$ satisfies the same constraint algebra (\ref{HtHt})-(\ref{HiHj}) or (\ref{HtHt_sm})-(\ref{HiHj_sm}). }}
\begin{\eq}
&& \nabla_i \left[N^2 ~ \widetilde{\Theta}^i|_{N \ra (u_t-{N_i \nabla^i N})}+2 (u_t-{N_i \nabla^i N}) N\widetilde{\Theta}^i   \right]
 \approx -\{\widetilde{\Xi} , {\overline{H}_C} \}
\no \\
&&~~~=\left(\f{\ka^2}{2} \right)^2 4 \xi \widehat{\la} (D-1) \nabla_i \left\{
 {N^2} \left[2N  \pi^{ij} \nabla_j \left[ \left(\xi R+ \f{\La D}{D-1} \right)N -\xi \nabla^2 N \right] \right. \right.\no \\
&&~~~~\left. \left.+~\xi \nabla^i \left[\pi^{jk} \left(R_{jk} N^2 -2 N \nabla_j \nabla_k N -\nabla_j N \nabla_k N \right) \right]
\right] \right\} {-\nabla_j \left( N^j \widetilde{\Xi} \right) }\no \\
&&~~~~+(\pi, \nabla_i \pi-\mbox{dependent terms}) +\mbox{(higher-derivative contributions)}.
\label{Phi_time_evol_final}
\end{\eq}
Here, it is important to note that the multipliers, $u_t,\nabla^i u_t, etc.$, have
generally non-vanishing coefficients in the left-hand side
so that (\ref{Phi_time_evol_final}) may provide the equation for determining the multiplier $u_t$. However, the similar equation for $u_i$ does not exist and  $u_i$ is still undetermined but this is just a reflection of the first-class nature of the constraint, $\pi^i\approx 0$, in (\ref{primary_cons}). This would now complete the Dirac's procedure for finding the complete set of constraints, though we would not try to solve for the explicit solution of the multiplier $u_t$. Then, the full set of constraints are given by $\chi_A \equiv ( \pi_N, {\cal H}, \widetilde{\Om}, \widetilde{\Xi} )\approx 0, \Ga_B \equiv (\pi^i, {\cal H}_i ) \approx 0$. Here,  the constraints
$\chi_A  \approx 0$ are the
{\it second-class} constraints with the constraint algebra,
\begin{\eq}
&&\{  \pi_N (x), {\cal H} (y) \} =0 , \no \\
&&\{   \pi_N (x), \widetilde{\Om} (y) \} = -2 \nabla^y_i \left( N \widetilde{C}^i(y) \de^D(x-y)\right)\approx 0, \no \\
&&\{  \pi_N (x), \widetilde{\Xi} (y) \} =\Delta (x-y), \no \\
&&\{ {\cal H} (x), {\cal H} (y) \} =C^i (x) \nabla^x_i \de^D (x-y)-C^i (y) \nabla^y_i \de^D (x-y)\approx 0 , \no \\
&&\{ {\cal H} (x),  \widetilde{\Omega} (y) \} \approx \{  \pi_N (x), \widetilde{\Xi} (y) \} ,~ etc.,
\label{const_Poisson_2_local_text}
\end{\eq}
whose determinant, $\mbox{det} \{\chi_A, \chi_B \} $, is generally non-vanishing,
\begin{\eq}
\mbox{det} \{\chi_A (x), \chi_B (y) \}&\approx &
\left( \mbox{det} \{ {\cal H} (x), \widetilde{\Om} (y) \} \right)^2
\left(\mbox{det} \{\widetilde{\Om} (x), {\cal H} (y) \} \right)^2
\no \\
& \approx & \left(\mbox{det} [ \Delta (x-y) \Delta (y-x) ] \right)^2,
\label{det_chi}
\end{\eq}
where
\beq
\Delta (x-y) &\equiv& -\nabla_{i}^{y} \left[  2 N \widetilde{\Theta}^i (y) ~\de^D(x-y)
 +N^2(y) \left( \f{\de \widetilde{\Theta}^i (y)}{\de N(x)}\right) \right] \no \\
 &=&-2 \widetilde{\Xi} ~\de^D(x-y)-  2 N^2 \widetilde{\Theta}^i (x) \nabla^x_i \left(\f{\de^D(x-y)}{N(x)} \right)-\nabla_i^y \left[ N^2(y) \left( \f{\de \widetilde{\Theta}^i (y)}{\de N(x)}\right) \right]
\eeq
with
\beq
&& \left( \f{\de \widetilde{\Theta}^i (y)}{\de N(x)} \right)
= -\left(\f{\ka^2}{2} \right)^2 \f{2 \widehat{\la}  {\cal V}'}{(\la D-1) \sqrt{g}} \left\{ \left[ (2 \la+1) g^{ij} \pi-2 (\la D-1) \pi^{ij}\right]
\left( \nabla_j \pi(y)+\pi^2 \nabla^i_y \right) \de^D (x-y) \right\} \no \\
&&~~-\left(\f{\ka^2}{2} \right) 2 \widehat{\la} (D-1)  \sqrt{g} {\cal V}' \nabla^i_y
\left[ \left(\xi R+\f{ \La D}{D-1}
 + \frac{D-n}{D-1} \alpha R^{n} \right) \de^D (x-y)  +\nabla^2 \left(\mathcal{V}^{\prime} \de^D (x-y) \right) \right] \no \\
&&~~-\left(\f{\ka^2}{2} \right) \left[ 2 \widehat{\la} \nabla_i \pi \{{\cal V}'(R)(y), {\cal H} (x) \}+
\{2 (\pi^{ij}-\widehat{\la} g^{ij} \pi ) \nabla_j {\cal V}'(R) (y), {\cal H}(x) \}\right].
\eeq

On the other hand, the constraints,
$\Ga_A \equiv (\pi^i, {\cal H}_i ) \approx 0$,
are the {\it first-class} constraints with the vanishing 
determinant,
$\mbox{det} (\{\Ga_A, \Ga_B \} )=0$.
Then, the resulting number of dynamical degrees of freedom in the ``configuration" space is given by
\begin{\eq}
s&=&\f{1}{2} \left( P-2 N_1-N_2\right) \no \\
&=&\f{1}{2} \left[ (D+1)(D+2)-2 \times 2 D -``4"\right] \no \\
&=&\f{1}{2} (D+1)(D-2),
\end{\eq}
where $P=(D+1)(D+2)$ is the number of canonical variables in the ``phase" space $(N, \pi_N, N_i, \pi^i,g_{ij}, \pi_{ij})$, $N_1=2D$ is the number of the first-class constraints $(\pi^i, {\cal H}_i)\approx 0$, and $N_2=``4"$ is the number of the second-class constraints,
$(\pi_N, {\cal H}, \widetilde{\Om}, \widetilde{\Xi} )\approx 0$. Note that, for Case
{\bf A}, 
the dynamical degrees of freedom are the same as that of GR (in arbitrary dimensions)
though the constraint structure is different \footnote{Recent constructions of, so called, ``minimally-modified" gravity theories \ci{Lin:2017,Carb:2018,Muko:2019,Gao:2019} may correspond to this case also.}: Actually, in GR, {\it i.e.,}
$\widehat{\la}=0,~ \al=0$ or the $\la$-deformed GR ($\la R$ model) with the condition, $\pi = 0$, we
have $N_1=2 (D+1), N_2=0$ so that the {\it 2 first-class} constraints,
$(\pi_N, {\cal H})\approx 0$, in GR or $\la$-deformed GR, transform into the {\it 4 second-class} constraints,
$( \pi_N, {\cal H}, \widetilde{\Om}, \widetilde{\Xi} )\approx 0$, in the Case {\bf A} of
full Ho\v{r}ava gravity,
with maintaining the same dynamical degrees of freedom $s$. This completes the
previous linear analysis in \ci{Park:0910b,Shin:1701}, but now
at the ``fully non-linear" level. (cf. \ci{Bell:1004,Deve:2019}).\\

{\bf B}. Case $\widetilde{C}^i \neq 0$: This is the more generic case where the conjugate momenta $\pi^{ij}$ and the (scalar) curvature $R$ are arbitrary, with the generic higher-derivative potential, ${\cal V}(R)$. In this case, (\ref{Om_dot}) does not yield new constraints but determines the multiplier $u^t$ generally \footnote{For some detailed discussion about the determination of the multiplier $u^t$, see \cite{Bell:1010} (see also \ci{Pons:2010} for an earlier discussion).} so that the Dirac's procedure may be completed, without further iterations. Then, in contrast to Case {\bf A}, there are the second-class constraints, $\widetilde{\chi}_A \equiv ( \pi_N, {\cal H}, \Om )\approx 0$, whose determinant
$\mbox{det}(\{\widetilde{\chi}_A, \widetilde{\chi}_B \})$ is non-vanishing,
generally,
\beq
\mbox{det} \{\widetilde{\chi}_A (x), \widetilde{\chi}_B (y) \}
&=&- \mbox{det} \{ \pi_N (x), \widetilde{\Om} (y) \}
~\mbox{det} \{ {\cal H} (x), {\cal H} (y) \} ~\mbox{det} \{  \widetilde{\Om} (x),\pi_N (y)\} \no \\
 & \approx & 4 ~ \mbox{det} \left( N \widetilde{C}^j (y) \nabla^y_j \de^D(x-y)\right)
 \mbox{det} \left( N \widetilde{C}^k (x) \nabla^x_k \de^D (x-y)\right) \no \\
  &&~ \times \mbox{det} \left( C^i (x) \nabla^x_i \de^D (x-y) \right)
-(x \lra y),
\eeq
whereas the first-class constraints, $\Ga_A \equiv (\pi^i, {\cal H}_i )$, are the same as in  Case {\bf A}. Hence, the resulting number of dynamical degrees of freedom is
\begin{\eq}
s&=&\f{1}{2} \left[ (D+1)(D+2)-2 \times 2 D -``3"\right] \no \\
&=&\f{1}{2} (D+1)(D-2)  +\f{1}{2},
\end{\eq}
with $N_1=2 D$ and $N_2=``3"$, which shows {\it one extra degree of freedom in phase space}, in addition to the usual $(D+1)(D-2)$ graviton (transverse traceless) modes in arbitrary (D+1)- dimensions \footnote{Here, we do not consider the extension terms which depend on the proper acceleration, $a_i=\pa_i N/N$, for simplicity. If we include these terms, in addition to the standard action (\ref{HL action}) \ci{Hora:0812,Hora}, the extra modes have ``two" phase space degrees of freedom, like the ordinary scalar fields \cite{Blas:0909,Bell:1106,Donn:2011,Gao:2014}, and this may become another different case, say, Case {\bf D}. Actually, this corresponds to an alternative formulation of the projectable case but now a smooth limit to the non-projectable case exists \cite{Blas:0909}.}. In particular, in (2+1)-dimensions, the extra mode is the only dynamical degree of freedom. This result supports the previous case-by-case results \ci{Bell:1010,Deve:2019} but in a more generic set-up with arbitrary dimensions and cosmological constant.

The usual increase of dynamical degrees of freedom with higher-time derivatives is the result of the competition between the increased canonical momenta for the higher-time derivative fields and their increased, associated new constraints \ci{Buch:1987,Buch:1991}. However, for Ho\v{r}ava gravity, there are no increased canonical momenta but exist only the increased {\it second-class} constraints: For Case {\bf A}, the increased second-class constraints are enough to preserve the dynamical degrees of freedom of GR, whereas for Case {\bf B}, they are not enough and one extra degree of freedom persists in the phase-space.

\section{Constraint analysis at the Conformal Point, $\la=1/D$: Case C}

The kinetic part of Ho\v{r}ava action (\ref{HL action}) can be written formally as,
\beq
S_K 
\equiv\left(\f{2}{\ka^2} \right) \int dt d^D x \sqrt{g} N  \left(K_{ij} G^{ijkl} K_{kl} \right),
\eeq
where $G^{ijkl} \equiv \de^{ijkl}-\la g^{ij} g^{kl}$ is the (generalized) DeWitt metric
\cite{DeWi,Hora:0812,Hora} \footnote{$\la_{Horava}=-\la_{DeWitt}/2$.}.
In the previous sections, we have considered the case, $\la \neq 1/D$, so that the DeWitt metric is not degenerated. Then the first-order action (\ref{HL action_1st}) can be obtained by considering the Legendre transformation of $S_K$,
\beq
S_K 
= \left(\f{\ka^2}{2} \right) \int dt d^D x \f{N}{2 \sqrt{g}}  \left( \pi^{ij} {\cal G}_{ijkl}\pi^{kl} \right),
\eeq
with the conjugate momenta, $\pi^{ij}=({2 \sqrt{g}}/{\kappa^2}) {G}^{ijkl}K_{kl}$, and the {\it inverse} DeWitt metric, ${\cal G}_{ijkl}=\de_{ijkl}-\la  g_{ij} g_{kl} /(D \la-1)$, satisfying $G^{ijmn}{\cal G}_{mnkl}={\cal G}_{klmn} G^{mnij}=\de^{ij}_{kl}$ \ci{Hora:0812,Hora}.

On the other hand, for $\la=1/D$, the DeWitt
metric is degenerated and we need to project out the non-degenerate
parts only when considering the appropriate inverse of the DeWitt metric. Actually,
using the fact that $\widehat{G}^{ijkl}\equiv G^{ijkl} |_{\la=1/D}$ has a null eigenvector $g_{ij}$,
\begin{\eq}
\widehat{G}^{ijkl} g_{ij}=0,
\end{\eq}
it is easy to see that its inverse, $\widehat{\cal G}_{ijkl}$, is given by
\begin{\eq}
\widehat{\cal G}_{ijkl}&=&\de_{ijkl}-\f{1}{D} g_{ij}g_{kl}, \no \\
\widehat{\cal G}_{ijkl} g^{ij} &=&0,~\widehat{\cal G}_{ijmn}
\widehat{G}^{mnkl}={\widetilde{\de}_{ij}}^{kl}
\end{\eq}
with the (projected) Kronecker-delta,
${\widehat{\de}_{ij}}^{kl}=\de_{ij}^{kl}-g_{ij} g^{kl}/D$,
satisfying ${\widehat{\de}_{ij}}^{kl}g^{ij}={\widehat{\de}_{ij}}^{kl}g_{kl}=0$
\ci{Park:0910b}. (See also \ci{Buch:1987,Buch:1991} for the corresponding analyses in
{\it Lorentz invariant} higher-curvature gravities.)

In the Hamiltonian formulation, the existence of a null eigenvector in $g_{ij}$ is
reflected in the primary constraint,
\begin{\eq}
\chi \equiv {\widehat{\pi}^i}_i \equiv g_{ij} \widehat{\pi}^{ij} \approx 0
\label{pi_constraint}
\end{\eq}
for the momenta,
\begin{\eq}
\widehat{\pi}^{ij} \equiv \f{\de S}{\de \dot{g}_{ij}} =\f{2
\sqrt{g}}{\kappa^2} \widehat{G}^{ijkl}K_{kl},
\end{\eq}
in addition to the usual primary constraints, $\Phi_{\mu}$, in (\ref{primary_cons}).

Then, one can find the canonical Hamiltonian, up to boundary terms,
\begin{\eq}
\label{Hc_conformal}
 \widehat{H}_C&=& \int_{\Si_t} d^{D}x  \left\{ N \widehat{\cal H}+N_i {\widehat{\cal H}}^i\right\}
\end{\eq}
with
\begin{\eq}
\widehat{\cal H}&\equiv&\frac{\kappa^2}{2 \sqrt{g}} \widehat{\pi}^{ij}\widehat{\pi}_{ij}  +\sqrt{g} {\cal V},
\label{H_t_conformal}
\\
\widehat{\cal H}^i &\equiv&-2 \nabla_j \widehat{\pi}^{ij}.
\label{H_i_conformal}
\end{\eq}

Now, the preservation of the additional primary constraint (\ref{pi_constraint}),
\beq
\dot{\chi}&=& \{ \chi, \widehat{H}_C\} \no \\
 &\equiv& \Psi \approx 0
\eeq
produces a new secondary constraint,
\beq
\Psi &\equiv& \left(\f{\ka^2}{2} \right) \left(\f{D}{2 } \right)\f{N}{\sqg} \widehat{\pi}_{mn} \widehat{\pi}^{mn} -N \sqrt{g} \left[ \left(\f{D}{2}\right) {\cal V}(R)-
R {\cal V}' (R)  \right] +\sqg (D-1) \nabla^2 \left(N {\cal V}'(R) \right) \no \\
&=&\f{D}{2} N \widehat{\cal H}-N \sqrt{g} \left[ D {\cal V}(R)-R {\cal V}' (R)  \right] +\sqg (D-1) \nabla^2 \left(N {\cal V}'(R) \right) \approx 0
\eeq
from (\ref{pi_trace_dot_App}), in addition to the usual (reduced) Hamiltonian and momentum constraints in  (\ref{secondary_const}),
\beq
\widehat{\cal H} \approx 0, ~\widehat{\cal H}^i \approx 0.
\eeq
The extended Hamiltonian is then given by
\begin{\eq}
\label{HE_conf}
 \widehat{H}_E&=&\widehat{H}_C + \int_{\Si_t} d^{D}x  ( u_{\mu} \Phi^{\mu} + v \chi)
\end{\eq}
with a new Lagrange multiplier $v$. The constraint algebra, (\ref{HtHt})-(\ref{HiHj}), are reduced to
\begin{\eq}
 \{ \widehat{\cal H} (x),\widehat{\cal H} (y) \}&=&\widehat{C}^i (x) \nabla^x_i \de^D (x-y)-\widehat{C}^i (y) \nabla^y_i \de^D (x-y), \label{HtHt_conf}\\
 \{ \widehat{\cal H} (x),\widehat{\cal H}_i (y) \}&=&-\widehat{\cal H} (y) \nabla^y_i
 \de^D (x-y), \label{HtHi_conf}\\
 \{ \widehat{\cal H}_i (x),\widehat{\cal H}_j (y) \}&=&\widehat{\cal H}_i  (y) \nabla^x_j \de^D (x-y)+\widehat{\cal H}_j  (x) \nabla^x_i \de^D (x-y), \label{HiHj_conf}
\end{\eq}
where
\begin{\eq}
\widehat{C}^i=\left(-\f{\ka^2}{2} \right) \left[\left( \widehat{\cal H}^i+2 \nabla^i \widehat{\pi} \right) {\cal V}' (R) +2 \left( \widehat{\pi}^{ij}-g^{ij} \widehat{\pi} \right) \nabla_j  {\cal V}' (R)  \right].
\label{Ci_1_conf}
\end{\eq}

Using the above reduced constraint algebra, one can find that
\begin{\eq}
\dot{\widehat{\cal H}}&=&\{\widehat{\cal H} , \widehat{H}_E \} \no \\
&=&\f{1}{N} \nabla_i (N^2 \widehat{C}^i) +\nabla_i (N^i \widehat{\cal H})
+\{\widehat{\cal H}, \left< v \pi \right> \}, \label{H_t_hat_evol}\\
\dot{\widehat{\cal H}_i}&=&\{\widehat{\cal H}_i , \widehat{H}_E \} \no \\
&=& \widehat{\cal H} \nabla_i N +\nabla_j (N^j \widehat{\cal H}_i)+\widehat{\cal H}_j \nabla_i N^j +\widehat{\pi} \nabla_i v  \approx 0,
\end{\eq}
where
\beq
\{\widehat{\cal H}, \left< v \pi \right> \}=\left[-\f{v}{N} \Psi +\sqrt{g}(D-1)\left( \f{v}{N} \nabla^2 \left(N \mathcal{V}^\prime (R) \right)-\mathcal{V}^\prime (R) \nabla^2 v \right)\right].
\label{73}
\eeq
For the potential, $-{\cal V}(R)\equiv\La+\xi R +\al R^n$, (\ref{73}) becomes
\beq
\{\widehat{\cal H}, \left< v \pi \right> \}=-\left[\f{v}{N} \Psi +\sqrt{g}(D-1)\left( -\f{v}{N} \nabla^2 \left(N  (\xi+\al n R^{n-1}) \right)+ \left(\xi+\al n R^{n-1}\right) \nabla^2 v \right) \right].
\label{73_b}
\eeq
Since we are considering the non-trivial case of $-{\cal V}' (R) \equiv \xi+\al n R^{n-1}\neq 0$, preserving the Hamiltonian constraint, $\widehat{\cal H}\approx 0$, {\it i.e.,} $\dot{\widehat{\cal H}} \approx 0$ in (\ref{H_t_hat_evol}), does not produce new constraints but determines the Lagrange multiplier $v$: For the $\la$-deformed GR ($\al=0$), where $\widehat{C}^i = 0$, (\ref{H_t_hat_evol}) and (\ref{73_b}) determine $v=N$.

On the other hand, for the preservation of the secondary constraint, $\widetilde{\Psi}\equiv \Psi_{\widehat{\cal H}\approx 0}$,
\beq
\widetilde{\Psi} \equiv N \sqrt{g} \left[ D \La +(D-1) \xi R +(D-n) \al R^n \right]+ {\sqg (D-1) \nabla^2 \left(N \mathcal{V}^\prime \right)},
\label{Psi_const}
\eeq
one can find that
\begin{\eq}
\dot{\widetilde{\Psi}}&=&\{\widetilde{\Psi} , \widehat{H}_E \} \no \\
&=&\{\widetilde{\Psi} , \widehat{H}_C \}+\{\widetilde{\Psi}, \left< u_t \pi_N\right> \}
+\{\widetilde{\Psi}, \left< v \widehat{\pi} \right> \} \label{Psi_hat_evol},
\end{\eq}
where \footnote{Useful relations for these computations are $\{\sqg (x), \pi (y)\}=({D}/{2}) \sqg (x)\de^D (x-y)$ and $\{ R(x), \pi (y) \}=-R (x) \de^D(x-y)-(D-1) \nabla^2_x \de^D(x-y)$ from (\ref{R_pi_Poisson}).}
\beq
&&\{\widetilde{\Psi}, \left< u_t \pi_N \right> \}=\f{u_t}{N} \widetilde{\Psi}   +\sqrt{g}(D-1)\left(  \nabla^2 \left(u_t \mathcal{V}^\prime \right)-\f{u_t}{N}  \nabla^2 \left(N\mathcal{V}^\prime \right) \right), \\
&&\{\widetilde{\Psi}, \left< v \pi \right> \}=v N \sqg \left[\f{D^2}{2} \La+\f{(D-1)(D-2)}{2} \xi R
+\f{(D-2 )(D-2 n)}{2} \al R^n\right] \no \\
&&~~~~~~~~+\sqg (D-1)  \left[ \f{(D-2)}{2} \nabla_i {\left(v \nabla^i \left( N \mathcal{V}^\prime \right) \right)-N \nabla^2 v \left( (D-1) \xi+ (D-n) \al n R^{n-1} \right) }\right] \no \\
&&~~~~~~~~+\al n(n-1)  (D-1) \sqg \nabla^2 \left[ R^{n-2} N \left( R v+(D-1) \nabla^2 v \right) \right].
\eeq
Now, (\ref{H_t_hat_evol}) and (\ref{Psi_hat_evol}) determine the multipliers $v$ and $u_t$,
respectively, so that one can finish the Dirac's procedure without generating further constraints.
Then, the complete set of constraints are
$\widehat{\chi}_A \equiv ( \pi_N, \widehat{\cal H}, \pi, \widetilde{\Psi} )\approx 0$ and
$\Ga_A \equiv (\pi^i, {\cal H}_i )\approx 0$. Here, the constraints,
$\widehat{\chi}_A \approx 0$, are
the {\it second-class} constraints with the algebra,
\begin{\eq}
&&\{ \pi_N (x), \widehat{\cal H} (y) \} =0 ,~~\{ \pi_N (x), \chi (y) \} = 0, \no \\
&&\{ \pi_N (x), \widetilde{\Psi} (y) \} = \widehat{\Delta} (x-y), \no \\
&&\{ \widehat{\cal H} (x), \widehat{\cal H} (y) \} =\widehat{C}^i (x) \nabla^x_i \de^D (x-y)-\widehat{C}^i (y) \nabla^y_i \de^D (x-y) , \no \\
&&\{ \widehat{\cal H} (x), \pi (y) \} \approx \{ \pi_N (x), \widetilde{\Psi} (y) \}
, etc.,
\label{const_Poisson_2_local_text}
\end{\eq}
whose determinant, $\mbox{det}(\{\widehat{\chi_A}, \widehat{\chi_B} \})$, is generally non-vanishing,
\begin{\eq}
\mbox{det} \{\widehat{\chi_A}, \widehat{\chi_B} \} &\approx &
\left( \mbox{det} \{ \widehat{\cal H} (x), \pi (y) \} \right)^2
\left( \mbox{det} \{\pi (x), \widehat{\cal H} (y) \}  \right)^2
\no \\
 &\approx& \left( \mbox{det} \left[ \widehat{\Delta} (x-y) \widehat{\Delta} (y-x) \right] \right)^2,
\label{det_chi}
\end{\eq}
where
\beq
\widehat{\Delta} (x-y) &\equiv& -\sqg \left[D \La+(D-1) \xi R +(D-n) \al R^n \right]\de^D(x-y) \no \\
&&+\sqg (D-1) \nabla^2_{y} \left[ ~\de^D(x-y) \left(\xi+\al n R^{n-1} \right)(y) \right].
\eeq
On the other hand, the constraints, $\Ga_A \equiv (\pi^i, {\cal H}_i )\approx 0$, are the first-class constraints, as in Cases {\bf A} and {\bf B}. So, the resulting number of dynamical degrees of freedom is the same as in Case {\bf A},
\begin{\eq}
s&=&\f{1}{2} \left( P-2 N_1-N_2\right) \no \\
&=&\f{1}{2} (D+1)(D-2),
\end{\eq}
which is the same as in GR. This provides the fully {\it non-perturbative} proof of the
previous perturbative analysis (see also \ci{Bell:2013} for an earlier work), which does not
show the extra degrees of freedom at the {\it linear} level \ci{Park:0910b}. There may exist
some similarities with Case {\bf A} due to the same physical degrees of freedom. Actually,
one can consider the maximal slicing condition, {\it i.e.,} $\pi =0$, for $\lambda$-deformed GR
(with an arbitrary $\lambda$), as an example satisfying the condition, $\widetilde{C}^i =0$ for
Case {\bf A}. But, an importance difference is that Case {\bf C} does not depend
whether $\widehat{C}^i=0$, {\it i.e.,} {\it commuting} Hamiltonian constraint
$\widehat{\cal H}$, or $\widehat{C}^i \neq 0$, {\it i.e.,} {\it non-commuting} Hamiltonian
constraint $\widehat{\cal H}$:
If we consider the maximal slicing, $\pi =0$, for Case {\bf B} with an arbitrary $\la \neq 1/D$, we have basically the same results as Case {\bf C} ! This implies that the constraint
 structure of Case {\bf C} and so its number of dynamical degrees of freedom do not depend on  spatially-higher-derivative terms in the potential, which are important for distinguishing Cases {\bf A} and {\bf B}. In other words, Case {\bf C} does not depend on the UV conformal symmetry for the Cotton square term, $C^{ij}C_{ij}$, in the Ho\v{r}ava's original potential \ci{Hora:0812,Hora} and this is essentially due to the kinematic origin of the constraint, $\chi=\widehat{\pi}\approx 0$, in (\ref{pi_constraint}) \footnote{This case corresponds to the $\dot{\beta}=0$ case in the Birkhoff's analysis of spherically symmetric system since $\pi \sim \dot{\beta}$ and the results are in a good agreement \ci{Deve:2019}.}.

\section{Examples}
In this section, we consider several exact solutions for our typical potential, $-{\cal V}(R) \equiv \xi R + \al R^n+\La$, as some explicit examples of the constraint analysis.

\subsection{Non-Rotating Black Holes in Arbitrary Dimensions}
For non-rotating, spherically symmetric black holes with $N_i \equiv 0$, the extrinsic curvature, $K_{ij}$, and the conjugate momentum, $\pi_{ij}$, vanish so that the Hamiltonian constraint becomes simply ${\cal H}=\sqrt{g} {\cal V}(R) \approx 0$. The general solution is `$R=constant$', whose constant value depends on the theory parameters, $\xi, \al, \La$, and $n$. If we consider, as an explicit example,  $D=3,~n=2$, {\it i.e.,} $z=2$ case in (3+1)-dimensions \ci{Kiri:2009} \footnote{This can be also the solution for $z=3$ case with the Cotton square term, $C_{ij}C^{ij}$, since the Cotton tensor vanishes, $C^{ij}\equiv \ep^{ikl} \nabla_k \left({R^j}_l-{\de^j}_l R  /4 \right)=0$, for the spherically symmetric cases.}, then the solution is given by
\beq
ds^2=-N^2(r) dt^2+\f{dr^2}{f(r)}+r^2 d \theta^2 +r^2 sin^2 \theta d \phi^2
\eeq
with
\beq
N^2&=&f=1-\f{2 M}{r} +\left(\f{\xi -\sqrt{\xi^2 -4 \al \La}}{12 \al} \right)r^2, \label{N_non-rotating} \\
R&=&\f{-\xi +\sqrt{\xi^2 -4 \al \La}}{2 \al}
\label{R_non-rotating} .
\eeq
In this case, even though there are higher-derivative contributions, we have $\widetilde{C}^i =0$, trivially, due to $\pi_{ij}=0$ in (\ref{Ci_3}). This would be a trivial example though its result is generally valid for arbitrary power $n$ and dimension $D$. If we consider the more general, spherically symmetric solutions with $N_r \neq 0$ (cf. \ci{Capa:2009}), it would be a more non-trivial example with the extrinsic curvature and conjugate momenta.

\subsection{Black String Solutions in (3+1)-Dimensions}
For stationary black strings in $(3+1)$-dimensions, the general ansatz is given by
\beq
ds^2=(-N^2+N_r N^r+N_\phi N^\phi ) dt^2+2 (N_r dr+N_\phi d \phi) dt+\f{dr^2}{f}+r^2 d \phi^2 +g dz^2,
\label{black_string}
\eeq
where all the metric functions, $N, N_i, f$, and $g$ depend on the radial coordinate $r$. For $z=2$ black string solution which satisfies the vanishing Cotton tensor, $C_{ij}=0$, for simplicity, a simple solution with $g=constant, N_{\phi}=0$, and $\la=1$, which is called BTZ-type black string, is given by (with $g \equiv 1$)
\beq
N^2&=&f=\eta r^2 -m, \\
N_r^2 &=&f^{-1} \left[ \de+\f{\ka^4 \mu^2}{64} \left(\eta^2-2 (\La_W +\om)-3 \La_W^2\right) \right],
\eeq
where $\eta, m$, $\de$ are integration constants \ci{Alie:2011} and the other parameters $\ka, \mu, \La_W, \om$ are introduced by the usual parametrization,
\beq
\xi \equiv \f{\ka^4 \mu^2 ( \La_W +\om)}{8 (1-3 \la)}, ~
\al \equiv \f{\ka^2 \mu^2 (1-4 \la)}{32 (1-3 \la)},~
 \La \equiv-\f{2 \ka^2 \mu^2 \La_W^2}{8 (1-3 \la)}.
\eeq
In this case, even though the extrinsic curvatures and conjugate momenta are non-vanishing, it has a constant curvature, $R \sim f'/r=2 \eta=constant$. Since this is the solution for $\la=1$, one can easily find that $\widetilde{C}^i=0$ in (\ref{Ci_3}) is trivially satisfied \footnote{This result is still valid with the angular shift vector, $N_{\phi}$.}: The first term in (\ref{Ci_3}) vanishes due to $\la=1$ and the second term vanishes due to $R=constant$. If we consider the more general solutions with $\la \neq 1$ or the $\la =1$ solution without the condition, $C_{ij}=0$, it would be more non-trivial examples with the {\it non-constant} extrinsic curvatures and conjugate momenta.

\subsection{Rotating Black Holes in (3+1)-Dimensions}
The exact solutions for rotating black holes in $(3+1)$-dimensions has not been found yet. However, for slowly rotating black holes, one can consider the ansatz \ci{Lee:2010,Alie:2010},
\beq
ds^2=-N^2(r) dt^2+\f{dr^2}{f(r)}+r^2 d \theta^2 +r^2 sin^2 \theta d \phi^2+2 a g(r) sin^2 \theta dt d \phi.
\eeq
At the linear order in the rotation parameter $a$, the solution of $g(r)$ for $n=2$ is given by
\beq
g(r)=\si r^2 +\f{\ga}{r}
\eeq
with the integration parameters, $\si, \ga$, and the same solution of $N^2=f(r)$ as in (\ref{N_non-rotating}). In this case, the extrinsic curvature has a non-vanishing component, $K_{r \phi}={\cal O}(a)$ but $K=0, \pi=0$, at the linear order in `$a$'. Since the curvature scalar is constant as in (\ref{R_non-rotating}), we will have the case $\widetilde{C}^i =0$ again and this result is valid for arbitrary power $n$ and dimensions $D$ \footnote{In \ci{Lee:2010,Alie:2010}, $\la=1, \La_W=0$ case for the IR-modified Ho\v{r}ava gravity was considered. But our result is generally valid for arbitrary $\la$ and $\La_W$. }. If we consider higher orders in the rotation parameter $a$ so that $K$ or $\pi$ is non-vanishing or $R$ is non-constant, it would be a more non-trivial example.

\subsection{Rotating Black Holes in (2+1)-Dimensions}
For stationary black holes in $(2+1)$-dimensions, the most general ansatz is given by
\beq
ds^2=(-N^2+N_r N^r+N_\phi N^\phi ) dt^2+2 (N_r dr+N_\phi d \phi) dt+\f{dr^2}{f}+r^2 d \phi^2,
\label{rot_black_hole_3}
\eeq
where all the metric functions, $N, N_i, f$, depend on the radial coordinate $r$, as in (\ref{black_string}), due to the symmetry property in $(2+1)$-dimensions. The general solution for the metric (\ref{rot_black_hole_3}) is not known yet. For the simple case, however, of $N_r=0$, one can find that the there is non-vanishing conjugate momentum, $\pi^{\phi r}=(f/2N) (N^{\phi})'$, whose trace is vanishing trivially, $\pi=0$, with the solution (for the $n=2$ case) \ci{Park:2012,Park:in prep}
\begin{\eq}
&&f=-{\cal M} +\f{b r^2}{2} \left[ 1- \sqrt{a+
\f{c}{r^4}}+\sqrt{\f{c}{r^4}}
ln\left(\sqrt{\f{c}{ar^4}}+\sqrt{1+\f{c}{ar^4}} \right)
\right], \no  \\
&&N^2/f=1/\left(1+\f{c}{ar^4}\right) , \no \\
&&N^{\phi}=-\f{{\cal J}}{2}\sqrt{\f{a}{c}}~ ln
\left[ \sqrt{\f{c}{a r^4}}+\sqrt{1+\f{c}{a r^4}} \right],
\label{f_Horava_3D_Kerr}
\end{\eq}
where
\begin{\eq}
a=1-\f{4 \al \Lambda}{\xi^2},~b=\f{\xi}{2 \al},~ c=\f{2 \al
{\cal J}^2}{\xi^2}. \label{abc}
\end{\eq}
In this case, even though $K=0,\pi=0$ trivially, the curvature scalar is non-constant,
\begin{\eq}
R=-\f{f'}{r}=-b \left(1-\sqrt{a+ \f{c}{r^4}} \right),
\label{R}
\end{\eq}
so that we have a non-vanishing component of $\widetilde{C}^i$,
\beq
\widetilde{C}^{\phi} \sim \al \pi^{\phi r} \partial_r R \neq 0.
\eeq
But, it easy to find that the constraint $\widetilde{\Om}\approx 0$ is satisfied again
\beq
\widetilde{\Om} \sim \pa_{\phi}\left(N^2 \widetilde{C}^{\phi} \right)=0,
\eeq
due to the spherical symmetry, {\it i.e.,} no $\phi$-dependence, in the solution. It is important to note that the non-vanishing $\widetilde{C}^i$ is the
genuine result of higher-derivative terms ($\al \neq 0$) with a rotation
($\pi^{\phi r} \neq 0$). In other words, if we turn off the higher-derivative term,
{\it i.e.,} $\al=0$, similarly to BTZ black hole in GR, the system is reduced to the case with
the vanishing $\widetilde{C}^i \sim \xi \widehat{\la} \nabla^i \pi =0$. Of course, this does not
correspond to Case {\bf B} but Case {\bf C}, due to the fact of $\pi=0$ for the solution
(\ref{f_Horava_3D_Kerr}), as can be seen by checking the  constraint,
$\widetilde{\Psi} \approx 0$, in (\ref{Psi_const}). However, if we consider the
$N_r \neq 0$ case, one obtains $\pi \neq 0$ generally so that it may correspond
to a ``genuine" case of Case {\bf B}. This example would show the importance of
higher-derivative terms for Case {\bf B}, where the extra scalar graviton mode is involved.

\section{Concluding Remarks}

We study the {\it full} constraint analysis of the {\it non-projectable} Ho\v{r}ava
gravity whose potential, ${\cal V}(R)$, is an arbitrary function of the
(intrinsic) Ricci scalar $R$ in arbitrary dimensions but without the extension terms which depend on the proper acceleration
$a_i$,
for simplicity.
We find that there are generally three distinct cases in this theory, depending on\\

(i) whether the Hamiltonian constraint, ${\cal H}\approx 0$, generates new (second-class) constraints (Cases {\bf A}, {\bf C}) or just fixes the associated Lagrange multipliers (Case {\bf B}), or\\

(ii) whether the { IR} Lorentz-deformation parameter $\la$ is at the conformal point, $\la=1/D$ (Case {\bf C}), or not (Cases {\bf A}, {\bf B}).\\

We find that, for Cases {\bf A} and {\bf C}, the dynamical degrees of freedom of Ho\v{r}ava gravity are the same as in GR, while for Case {\bf B}, there is {\it one extra phase-space degree of freedom}, representing an extra (odd) scalar graviton mode. This would achieve the dynamical consistency
of a restricted model at the {\it fully non-linear} level and be positive in resolving the long-standing debates about the extra graviton
modes of the Ho\v{r}ava gravity.
Several further remarks about other challenging problems are in order.\\

1.
We have obtained the new ``extended" constraint algebra for the Hamiltonian and momentum constraints, (\ref{HtHt})-(\ref{HiHj}) ((\ref{HtHt_conf})-(\ref{HiHj_conf}) for Case {\bf C}) or (\ref{HtHt_sm})-(\ref{HiHj_sm}), for the {\it non-projectable} Ho\v{r}ava gravity whose potential is an arbitrary function of the {\it Ricci} scalar $R$. The structure of the newly obtained ``extended" constraints algebra seems to be generic to Ho\v{r}ava gravity itself, analogous to that of general higher-curvature gravities \ci{Buch:1987,Buch:1991}, where $C^i$ becomes the momentum constraint ${\cal H}^i$ with higher-curvature corrections. An important difference is that the momentum constraint ${\cal H}^i$ in Ho\v{r}ava is the same form as in GR with no higher-derivative corrections and satisfies the same algebra as in GR \cite{DeWi} but the full algebra with the Hamiltonian constraint, ${\cal H}$, is {\it not closed}, whereas the momentum constraints, ${\cal H}^i \equiv C^i$, in generic higher-curvature gravities satisfy exactly the same {\it closed} algebra as in GR, ``{\it as has been argued generically in \ci{Teit:1972}}", even with the generic higher-curvature terms which include Riemann tensors also \ci{Buch:1987,Buch:1991,Deru:2009}. We suspect that a similar general argument in Ho\v{r}ava gravity exists also so that the extended constraint algebra reflects the generic space-time structure of our Lorentz violating gravities with {\it Ricci} and {\it Riemann} tensors, $R_{ij}, R_{ijkl}$, {\it etc.}, {\it i.e.,} with the potential, ${\cal V}(R,R_{ij}, R_{ijkl}, {\cdots})$ \footnote{This implies the {\it generic} absence of the third and second-derivative terms in (2.24) of \ci{Li}, which is the case of $\la=1$ and ${\cal V} \sim C_{ij} C^{ij}$.}, and its general proof would be a challenging problem.\\

2. Our constraint analysis shows the dynamical degrees of freedom in a restricted model of the Ho\v{r}ava gravity at the {\it fully non-linear} level. Comparison to the previous linear perturbation analyses
\ci{Park:0910a,Park:0910b,Shin:1701}, which do {\it not} show the extra degrees of freedom,
implies that the extra modes would be the {\it genuine consequence of non-linear
effect (around the {\it homogenous} background)} with Lorentz-violating higher-derivative
terms \ci{Deve:2019}: For a direct proof in the $n=2$ constraint algebra, see \ci{Bell:1010}.
On the other hand, it has been also argued that the extra mode may appear when considering
(even linear) perturbations around the spatially-{\it inhomogeneous} and
{\it time-dependent} background \ci{Blas:0906}, even with the IR Lorentz-violating terms only,
as anticipated from the usual connection between non-linear perturbations for a {\it homogenous} background and linear perturbations for a corresponding {\it inhomogeneous} background \footnote{MIP thank K. Koyama and A. E. Gumrukcuoglu for discussion about this matter.}. This may be in contrast to the Birkhoff's theorem analysis in {\it spherically symmetric, vacuum} configuration which does {\it not} show time-dependent solutions, representing the ``extra gravitational" modes \ci{Deve:2019}. However, this might be due to its high symmetry, {\it i.e.,} lower inhomogeneity, and there might exist still some possibilities for higher inhomogeneities: This might be consistent with a related analysis in \ci{Bell:1010} which does not show the extra mode only up to the first order of inhomogeneity, $L^{-1}$. So, a direct proof of the argument of \ci{Blas:0906} in the generic constraint algebra would be an interesting open problem. The role of non-linear, UV scalar graviton in Big Bang cosmology and gravitational wave physics for compact objects with strong gravities, like black holes and neutron stars, would be also a challenging problem.\\

3. In the literature, there have been claims of inconsistencies of the Ho\v{r}ava gravity, in its
original form. It seems that some originate from the limited analysis, like linear approximations
\ci{Blas:0909}, or incomplete constraint analysis \ci{Li,Henn}.
Here, we consider the second case \footnote{ The first case has
been discussed in several places, like \ci{Bell:1010} ({\it footnote 3}) and \ci{Li} ({\it Note added}). }, especially about the claim in \ci{Henn}, which seems to be the most rigorous criticism against the (non-projectable) Ho\v{r}ava gravity. The
basic claim in \ci{Henn} is that ``$N=0$ is the only possible solution" for the constraint
(4.1), $\nabla_i (N^2 \nabla^i \pi)\approx 0$, which corresponds to our constraint
(\ref{Omega_const}), $\widetilde{\Om}\equiv \nabla_i (N^2 \widetilde{C}^i)\approx 0$, for the
asymptotically flat ($\Lambda=0$) and $\la$-deformed GR,
otherwise $N$ blows up at infinity, generically \footnote{This result corresponds to exactly what has been argued in \cite{Blas:0906}, which shows ``instabilities for perturbations around a non-vanishing $K$ background", though given at the linear level.}. But, the elaborate analysis in \ci{Henn} is just another proof of an adequacy of the condition, $\pi=0$, in that example, which is the only remaining solution for the constraint (4.1) in \ci{Henn} or (\ref{Omega_const}) in this paper, as has been argued also in \ci{Bell:1004,Bell:1010}. Actually, contrary to the argument in \ci{Henn}, the perturbations from $\pi=0$ are not arbitrary but restricted by another constraint (\ref{Phi_constraint}), $\widetilde{\Xi} \approx 0$ in this paper (or (20) in \ci{Bell:1004}), when $\la \neq 1$. Moreover, when the metric is not asymptotically flat ($\Lambda \neq 0$), the argument of \ci{Bell:1004} (and possibly of \ci{Henn} or \ci{Blas:0906} also) would not be valid generally, as can be seen in the general, (2+1)-dimensional solution, (\ref{rot_black_hole_3}) with $N_r \neq 0$, which shows the non-vanishing $\pi$ and $N$. \\

4. With the full set of constraints, we can now consider Feynman's path-integral for the S-matrix elements \ci{Fadd:1969,Senj:1976}, whose Hamiltonian expression for Case {\bf A}, for example, is given by
\beq
{Z}&=&\int {\cal D} g_{ij} {\cal D} \pi^{ij} {\cal D} N {\cal D} N_{i} \de ({\cal H}_i ) \de (\gamma_j) \de({\cal H}) \de(\widetilde{\Om})~ \int {\cal D} c_j {\cal D} \bar{c}_i~ exp \left\{i \int dt d^D x  ~ \bar{c}_i \left[ \{ {\cal H}_i, \gamma_j \} |_{\pi^i \approx 0} \right]  c_j  \right\} \no \\
&& \times \int {\cal D} b {\cal D} \bar{b} ~ exp \left\{i \int dt d^D x  ~ \bar{b} \left[ \{ {\cal H}, \widetilde{\Om} \} \{\widetilde{\Om}, {\cal H} \}|_{\pi_N \approx 0} \right]  b +{i} S /\hbar\right\},
\label{path_int}
\eeq
where we have integrated out for the primary constraints $\pi_N, \pi^i \approx 0$ with the Faddeev-Popov's
anti-commuting fields, $c_j, \bar{c}_i, b, \bar{b}$, the gauge-fixing conditions, $\gamma_j=0$, for the first-class constraint,
${\cal H}_i\approx 0$, and the determinant factor
for the second-class constraints,
$\left( \mbox{det} \{ \chi_A, \chi_B \} \right)^{1/2} \approx
 \mbox{det} \left( \{ {\cal H}, \widetilde{\Om} \}
 \{\widetilde{\Om}, {\cal H} \} \right)$.
One can obtain similarly the path integral for Cases {\bf B} and {\bf C} also. It would be desirable to study the renormalizability for the non-projectable Ho\v{r}ava gravity, based on the above S-matrix elements, beyond the recent proof of (perturbative) renomalizability for the projetable cases  \ci{Barv:2015,Barv:2017,Barv:2019} \footnote{For an earlier work on non-projectable case in (2+1)-dimensions, see also \ci{Bell:2016} (Case {\bf C}).}.\\

5. In the canonical quantization with the second-class constraints, we need to compute the Dirac brackets \ci{Dira} \footnote{{After this paper appeared in the archive, we became aware of \ci{Das:2011} in which the Dirac brackets were computed for the linearized Ho\v{r}ava gravity. We thank S. Ghosh for informing us about his work.}}, whose definition for Case {\bf A}, for example, is given by, for any variable $P, Q$,
\beq
\{ P, Q\}^* \equiv \{P, Q\}- \int d^D z \int d^D w ~\{P, \chi_A (z) \} C^{-1}_{AB}(z,w)
\{\chi_B (w), Q \},
\eeq
where $C^{-1}_{AB}$ is defined as $\int d^D z~ C^{-1}_{AB} (x,z) C_{BC} (z,y)=\int d^D z ~C_{CB} (y,z) C^{-1}_{BA} (z,x)=\de_{AC} \de^D (x-y)$ for the Poisson brackets of the second-class constraints $C_{AB} \equiv \{ \chi_A, \chi_B \}$,
and given by
\[ C^{-1}_{AB}(x,z)=\left( \begin{array}{cccc}
0 &-\De^{-1}(z-x) &0 & 0 \\
\De^{-1}(x-z) & 0& 0& 0 \\
0& 0& 0& -\De^{-1}(z-x)\\
0& 0& \De^{-1}(x-z) &0
\end{array} \right) \]
with $\De^{-1}$, defined by $\int d^D z~ \De^{-1} (x-z) \De (z-y)=\int d^D z ~\De (y-z) \De^{-1} (z-x)=\de^D (x-y)$. The bracket satisfies $\{ \chi_A, Q \}^*=0$ for any variable $Q$ so that the second-class constraints, $\chi_A \approx 0$, can be imposed consistently in the Hamiltonian dynamics, {\it i.e.}, $\{ {\cal H}, {\cal H} \}^* =0, \{ {\cal H}, {\cal H}_i \}^* =0, etc.$
\footnote{This algebra looks like that of the ``ultra-local" theory of gravity
 \ci{Teit:1981,Hora:0812,Hora}.
}, which corresponds to the delta-function insertions for the constraints, $\chi_A \approx 0$, in the path integral, (\ref{path_int}). If we consider the gauge-fixing conditions as in the path integral formalism, we can obtain further corrections to the Dirac brackets. Then the Dirac's quantization rule is given by $[ \widehat{P}, \widehat{Q} ] \equiv (i/\hbar) \{ P, Q \}^*$ for the quantum operators, $\widehat{P}$ and $\widehat{Q}$, corresponding to classical variables $P$ and $Q$, respectively, with the``appropriate" operator orderings. One can consider also Cases {\bf B} and {\bf C} similarly, but it would be more involved for the former case. \\

{\it Note added}: After finishing this paper, a related paper \ci{Bell:2019} appeared which is
overlapping with ours for $D=2$ case. But due to the (full) $a_i$ extensions in \ci{Bell:2019}, it shows a different constraint structure, as noted in our footnote No. 9.

\section*{Acknowledgments}
This work was supported by Basic Science Research Program through the National Research Foundation of Korea (NRF) funded by the Ministry of Education, Science and Technology {(2016R1A2B401304, 2020R1A2C1010372, 2020R1A6A1A03047877)}. DOD was also supported by the National Natural Science Foundation of China under Grant No. 11875136 and the Major Program of the National Natural Science Foundation of China under Grant No. 11690021.

\appendix

\begin{section}
{Computing $\{ {\cal H}, {\cal H} \}, \{ {\cal H}, {\cal H}_i \}$, and $\{ {\cal H}_i, {\cal H}_j \}$}
\end{section}

In this Appendix, we compute the constraint algebra, $\{ {\cal H}, {\cal H} \}, \{ {\cal H}, {\cal H}_i \}$, and $\{ {\cal H}_i, {\cal H}_j \}$ in (\ref{HtHt})-(\ref{HiHj}). To this ends, it is useful to consider the variations of the smeared constraints,
$\left< \eta {\cal H} \right> \equiv \int d^D x~ \eta {\cal H}\approx 0$, {\it etc.}, with the smearing functions, $\eta$, $\eta_i$ (neglecting the boundary terms),
\begin{\eq}
\de \left< \eta {\cal H} \right>&=&\left< A^{kl} \de g_{kl} +B_{kl} \de \pi^{kl}\right>,
\label{A1}
\\
\de \left< \eta^i {\cal H}_i \right>&=&\left< C^{kl} \de g_{kl} +D_{kl} \de \pi^{kl}\right>.
\label{A2}
\end{\eq}
After long computations, one can obtain the coefficients, $A, B, C, D$ as follows:
\begin{\eq}
&&A^{kl} \equiv \f{\de \left< \eta {\cal H} \right>}{\de g_{kl}}
=\left( \frac{\kappa^2}{2} \right) \f{\eta}{ \sqrt{g}} \left[-\f{1}{2} g^{kl} \left( \pi_{mn} \pi^{mn}-\widetilde{\la} \pi^2 \right)
+2 \left( \pi^{km} \pi^l_m-\widetilde{\la} \pi \pi^{kl} \right) \right] \no \\
 &&~~~~+\eta \sqrt{g} \left[  \f{ 1}{2} g^{kl} {\cal V}(R)-R^{kl} {\cal V}'(R) \right]
 +\sqrt{g} \left[ \nabla^k \nabla^l \left( \eta {\cal V}'(R) \right)-g^{kl} \nabla_m \nabla^m \left( \eta {\cal V}'(R)\right) \right],
 \label{A3}\\
&&B_{kl} \equiv \f{\de \left< \eta {\cal H} \right>}{\de \pi^{kl}}
=\left( \frac{\kappa^2}{2} \right) \f{2 \eta}{ \sqrt{g}} \left(\pi_{kl}-\widetilde{\la}  g_{kl} \pi \right),
\label{A4}\\
&&C^{kl} \equiv\f{\de \left< \eta^i {\cal H}_i \right>}{\de g_{kl}}
= -\nabla_m \left(\eta^m \pi^{kl} \right) +\left(\nabla_m \eta^k\right) \pi^{lm}+\left(\nabla_m \eta^l\right) \pi^{km},
\label{A5} \\
&&D_{kl} \equiv \f{\de \left< \eta^i {\cal H}_i \right>}{\de \pi^{kl}}
=\nabla_k \eta_l + \nabla_l \eta_k.
\label{A6}
\end{\eq}
As a byproduct, one can also obtain a useful formula,
\beq
\{ R(x), \pi^{kl} (y) \}=\f{\de R (x)}{\de g_{kl} (y)}=-R^{kl}(x) \de^D (x-y)
+\nabla^k_x \nabla^l_x \de^D (x-y)-g^{kl} \nabla^2_x \de^D (x-y).
\label{R_pi_Poisson}
\eeq

Then, after some manipulations, one can find the Poisson bracket algebras for the smeared constraints as follows:
\begin{\eq}
 \{ \left<\eta {\cal H} \right>,\left<\zeta {\cal H} \right> \}
 &=&\int d^D z \left[ \f{\de \left< \eta {\cal H} \right>}{\de g_{kl} (z)} \f{\de \left< \zeta {\cal H} \right>}{\de \pi^{kl} (z)}- \f{\de \left< \eta {\cal H} \right>}{\de \pi^{kl}(z)}  \f{\de \left< \zeta {\cal H} \right>}{\de g_{kl} (z)}   \right] \no \\
 &=&\left< \left( \eta \nabla_i \zeta-\zeta \nabla_i \eta \right) C^i  \right> , \\
 \{ \left< \eta{\cal H} \right>,\left< \zeta^i {\cal H}_i\right>\}
 &=&\int d^D z \left[ \f{\de \left< \eta {\cal H} \right>}{\de g_{kl} (z)} \f{\de \left< \zeta^i {\cal H}_i \right>}{\de \pi^{kl} (z)}- \f{\de \left< \eta {\cal H} \right>}{\de \pi^{kl}(z)}  \f{\de \left< \zeta^i {\cal H}_i \right>}{\de g_{kl} (z)}   \right] \no \\
 &=&-\left< ~  \zeta^i \nabla_i \eta {\cal H} \right> , \\
 \{ \left< \zeta^i {\cal H}_i \right>,\left< \zeta^j {\cal H}_j \right> \}
 &=&\int d^D z \left[ \f{\de \left< \zeta^i {\cal H}_i \right>}{\de g_{kl} (z)} \f{\de \left< \zeta^j {\cal H}_j \right>}{\de \pi^{kl} (z)}- \f{\de \left< \zeta^i {\cal H}_i \right>}{\de \pi^{kl}(z)}  \f{\de \left< \zeta^j {\cal H}_j \right>}{\de g_{kl} (z)}   \right] \no \\
 &=&\left< \left( \zeta^i \nabla_i \zeta^j-\zeta^i \nabla_i \zeta^j \right) {\cal H}_j \right> ,
\end{\eq}
where
\begin{\eq}
C^i=\left(-\f{\ka^2}{2} \right) \left[\left({\cal H}^i + 2 \widehat{\la} \nabla^i \pi \right) {\cal V}' (R) +2 \left(\pi^{ij}-\widehat{\la} g^{ij} \pi \right) \nabla_j  {\cal V}' (R)  \right].
\label{Ci_1_Append}
\end{\eq}
Now, one can easily check that the Poisson algebra for the local constraints are given by (\ref{HtHt})-(\ref{HiHj}):
\begin{\eq}
 \{ {\cal H} (x),{\cal H} (y) \}&=&C^i (x) \nabla^x_i \de^D (x-y)-C^i (y) \nabla^y_i \de^D (x-y), \label{HtHt_Append}\\
 \{ {\cal H} (x),{\cal H}_i (y) \}&=&-{\cal H} (y) \nabla^y_i
 \de^D (x-y), \label{HtHi_Append}\\
 \{ {\cal H}_i (x),{\cal H}_j (y) \}&=&{\cal H}_i  (y) \nabla^x_j \de^D (x-y)+{\cal H}_j  (x) \nabla^x_i \de^D (x-y). \label{HiHj_Append}
\end{\eq}

Moreover, from (\ref{A1})-(\ref{A6}), one can easily obtain the dynamical equations of motion (\ref{g_dot})-(\ref{pi_dot}) as follows,
\begin{\eq}
\dot{g}_{ij}&=&\{ g_{ij}, \left<N {\cal H}+N_i {\cal H}^i \right> \} = \f{\de\left<N {\cal H} \right>}{\de \pi^{ij}} +\f{\de\left<N_i {\cal H}^i \right>}{\de \pi^{ij}} \no \\
&=&\left( \frac{\kappa^2}{2} \right) \f{2 N}{ \sqrt{g}} \left(\pi_{ij}-\widetilde{\la}  g_{ij} \pi \right) + \nabla_i N_j + \nabla_j N_i, \label{g_dot_App}\\
\dot{\pi}^{ij}&=&\{ \pi^{ij}, \left<N {\cal H}+N_i {\cal H}^i \right>  \}= -\f{\de \left<N {\cal H} \right>}{\de g_{ij}} -\f{\de\left<N_i {\cal H}^i \right>}{\de g_{ij}} \no \\
&=&\left( \frac{\kappa^2}{2} \right) \f{ N}{ \sqrt{g}} \left[\f{1}{2} g^{ij} \left( \pi_{mn} \pi^{mn}-\widetilde{\la} \pi^2 \right)
-2 \left( \pi^{im} \pi^j_m-\widetilde{\la} \pi \pi^{ij} \right) \right] \no \\
 &&-N \sqrt{g} \left[  \f{ 1}{2} g^{ij} {\cal V}(R)-R^{ij} {\cal V}'(R) \right]
 -\sqrt{g} \left[ \nabla^i \nabla^j \left( N {\cal V}'(R) \right)-g^{ij} \nabla_m \nabla^m \left( N {\cal V}'(R)\right) \right] \no \\
&& + \nabla_m \left(N^m \pi^{ij} \right) -\left(\nabla_m N^i\right) \pi^{jm}-\left(\nabla_m N^j\right) \pi^{im}. \label{pi_dot_App}
\end{\eq}
Combining (\ref{g_dot_App}) and (\ref{pi_dot_App}), one can also obtain the dynamical equation for $\sqrt{g}$ and the momentum trace $\pi$ as follows,
\begin{\eq}
\dot{\sqrt{g}}&=&\left( \frac{\kappa^2}{2} \right)\f{N \pi}{1-D \la} +2 \nabla_i N^i, \\
\dot{\pi}&=&\left( \frac{\kappa^2}{2} \right) \left(\f{D}{2}\right) \f{ N }{ \sqrt{g}} \left( \pi_{mn} \pi^{mn}-\widetilde{\la} \pi^2 \right)+ \nabla_m \left(N^m \pi \right) \no \\
&&-N \sqrt{g} \left[  \left(\f{ D}{2} \right) {\cal V}(R)-R {\cal V}'(R) \right]
 +\sqrt{g}(D-1) \nabla^2 \left( N {\cal V}'(R) \right) \no \\
&=& \f{D}{2} N {\cal H} -N \sqrt{g} \left[ D   {\cal V}(R)-R {\cal V}'(R) \right]
 +\sqrt{g}(D-1) \nabla^2 \left( N {\cal V}'(R) \right) + \nabla_m \left(N^m \pi \right).
\label{pi_trace_dot_App}
\end{\eq}

\begin{section}
{More Details of computations in Sec. III A }

In this Appendix, we present some more details of the computations in Sec. III {\bf A}.

First, we consider the variations of the smeared constraints, $\left< \eta \widetilde{\Omega} \right> \equiv \int d^D x~ \eta \widetilde{\Omega}$ and $\left< \zeta \widetilde{\Xi} \right> \equiv \int d^D x~ \zeta \widetilde{\Xi}$  with the smearing functions $\eta$ and $\zeta$,
\begin{\eq}
\de \left< \eta \widetilde{\Omega} \right>&=&\left< E^{mn} \de g_{mn} +F_{mn} \de \pi^{mn}\right>,
\label{B1}
\\
\de \left<  \zeta \widetilde{\Xi} \right>&=&\left< G^{mn} \de g_{mn} +H_{mn} \de \pi^{mn}\right>.
\label{B2}
\end{\eq}
After quite tedious computations, we obtain the coefficients, $E, F, G, H$ as follows
\footnote{Due to the messy expressions,
we have used ``{\it xAct}" for cross-checking our computations.}:
\begin{\eq}
E^{mn} &\equiv& \f{\de \left< \eta \widetilde{\Omega} \right>}{\de g_{mn}}
=-\left(\frac{\kappa^2}{2}\right)2 \widehat{\lambda} {\cal V}'(R) \left[\pi^{mn}\nabla_{i}(N^{2}\nabla^{i}\eta)
-\frac{1}{2}g^{mn}\nabla_{i}(N^{2}\pi\nabla^{i}\eta)
+N^{2}\nabla^{(m}\pi\nabla^{n)}\eta\right] \no \\
&&- \left(\f{\ka^2}{2} \right)
2 \widehat{\la}  \left< \eta \nabla_i \left[ N^2 \nabla^i \pi \f{\de {\cal V}' (R)}{\de g_{mn}}
+ N^2 \left(\pi^{ij}-\widehat{\la} g^{ij} \pi \right) \f{\de  \nabla_j  {\cal V}' (R) }{\de g_{mn}}  \right. \right.\no \\
&&\left. \left. -N^2  \left( {\de^{ij}}_{mn} \pi +g^{ij} \pi^{mn} \right)  \nabla_j  {\cal V}' (R)  \right]
\right>,
 \label{B3}\\
F_{mn} &\equiv& \f{\de \left< \eta \widetilde{\Omega} \right>}{\de \pi^{mn}}
=-\left(\frac{\kappa^2}{2}\right)2 \widehat{\lambda}  {\cal V}'(R) \nabla_{i}\left[g_{mn}N^{2}\nabla^{i}\eta\right] \no \\
&&- \left(\f{\ka^2}{2} \right) 2
\left< \eta \nabla_i \left[ N^2  \left( {\de^{ij}}_{mn}-\widehat{\la}g^{ij}g_{mn} \right)  \nabla_j
 {\cal V}' (R)  \right] \right>,
\label{B4}\\
G^{mn} &\equiv&\f{\de \left< \zeta \widetilde{\Xi} \right>}{\de g_{mn}}
=\left(\frac{\kappa^{2}}{2}\right)2(D-1)\xi^{2}\widehat{\lambda}\sqrt{g}\Bigg\{ \frac{1}{2}g^{mn}N^{2}\nabla_{i}\zeta (\nabla^{i}\nabla^{j}\nabla_{j}N-N\nabla^{i}R-R\nabla^{i}N )\nonumber\\
&&+N^{2}\Big[\nabla_{i}(NR^{mn})\nabla^{i}\zeta-\nabla^{(m}\nabla_{i}\nabla^{i}N\,\nabla^{n)}\zeta+\nabla^{(m}(NR)\nabla^{n)}\zeta-\nabla_{i}\nabla^{m}\nabla^{n}N\nabla^{i}\zeta\Big]\nonumber\\
&&+\frac{\Lambda D}{2\xi(D-1)}\Big[2N^2\nabla^{(m}N\nabla^{n )} \zeta-g^{mn}N^{2}\nabla^{i}N\nabla_{i}\zeta\Big]+\nabla_{i}\left[N^{2}(X^{imnkl}{}_{l}\nabla_{j}\nabla_{k}N\,\nabla^{j}\zeta \right. \no \\
&&\left. +2 X^{imnkjl}\nabla_{l}\nabla_{k}N\,\nabla_{j}\zeta-NR^{mn}\nabla^{i}\zeta)\right]
+\nabla_{k}\nabla_{i}\left[N^{2}(X^{kmnl}{}_{l}{}^{i} -X^{kmnil}{}_{l})\nabla_{j}N\nabla^{j}\zeta \right. \no \\
&&\left.-N^{2}X^{kmnjl}{}_{l}\nabla_{j}N\nabla^{i}\zeta\right]+\nabla_{k}\nabla_{j}\nabla_{i}\Big[N^{3}\nabla^{i}\zeta(X^{kmnjl}{}_{l}-X^{kmnl}{}_{l}{}^{j})\Big]\Bigg\}\no \\
&&+\left(\frac{\kappa^2}{2}\right)^2 \frac{\xi\widehat{\lambda}}{(\lambda D-1)\sqrt{g}}
\Bigg\{(2\lambda+1)\left[2\pi^{mn}\pi\nabla_{i}(N^{3}\nabla^{i}\zeta)-g^{mn}\nabla_{i}(N^{3}\pi^{2}\nabla^{i}\zeta) \right. \no \\
&&~~~~~\left.+g^{mn}N^{3}\pi\nabla^{i}\pi\nabla_{i}\zeta
+2N^{3}\pi\nabla^{(m}\pi\nabla^{n)}\zeta\right]+N^{2}\left[g^{mn}\pi^{2}\nabla_{i}N\nabla^{i}\zeta+2\pi^{2}\nabla^{(m}N\nabla^{n)}\zeta \right. \no \\ &&~~~~~\left.-4\pi^{mn}\pi\nabla_{i}N\nabla^{i}\zeta\right]+4(\lambda D-1)\left[\left(\frac{1}{2}g^{mn}\pi^{ij}\pi-\pi^{mn}\pi^{ij}\right)\nabla_{j}(N^{3}\nabla_{i}\zeta)\right]\Bigg\}
\no \\
&&+\mbox{(higher-derivative contributions)},
\label{B5} \\
H_{mn} &\equiv& \f{\de \left<\zeta \widetilde{\Xi} \right>}{\de \pi^{mn}}
=\left(\frac{\kappa^{2}}{2}\right)^{2}\frac{ 4\xi}{\sqrt{g}}\widehat{\lambda}\,\Big\{N^{3}\nabla_{(m}\pi\nabla_{n)}\zeta-g_{mn}\nabla_{i}(N^{3}\pi^{ij}\nabla_{j}\zeta) \no \\
&&~~~~~+\frac{N^{2}g_{mn}}{2(\lambda D-1)}\left[(2\lambda+1)N\pi\nabla_{i}\nabla^{i}\zeta
+(6\lambda+1)\pi\nabla_{i}N\nabla^{i}\zeta\right]\Big\} \no \\
&&~~~~~+\mbox{(higher-derivative contributions)},
\label{B6}
\end{\eq}
where $f_{(ij)} \equiv (f_{ij}+f_{ji})/2$, $-{\cal V}'(R) \equiv \left(\xi +\al n R^{n-1} \right)$, and
\begin{\eq}
-\f{\de {\cal V}' (R) (y)}{\de g_{mn}(x)}
&=& \al n (n-1) R^{n-2} \f{\de R (y)}{\de g_{mn}(x)} \no\\
&=&  -\al n (n-1) R^{n-2} \left( R^{mn} \de^D (x-y)+A^{mnpq}{}_{kl}g^{kl}\nabla_p \nabla_q \de^D (x-y) \right) \no \\
&=&-\al n (n-1) R^{n-2} \left[ R^{mn} \de^D (x-y){-(\nabla^{n}\nabla^{m}-g^{mn}\nabla^2)}\de^D (x-y) \right],\\
A^{mnpq}{}_{kl}&\equiv& g^{pi} X^{qmn}{}_{ikl}-g^{ij} X^{qmn}{}_{ij(k}\delta^{p}{}_{ l)}, \no \\
 X^{mij}{}_{nkl}&\equiv& \frac{1}{2}\left[g^{m}{}_{k}g^{(i}{}_{l}g^{j)}{}_{n} +g^{m}{}_{l}g^{(i}{}_{n}g^{j)}{}_{k} -g^{m}{}_{n}g^{(i}{}_{k}g^{j)}{}_{l} \right].
\end{\eq} \no

Then, after some manipulations one can find the Poisson bracket algebras as follows:
\begin{\eq}
 &&\{ \left<\eta\widetilde{\Omega}\right>,\left<\zeta\widetilde{\Omega}\right> \}
 =\left(\frac{\kappa^2}{2}\right)  \widehat{\lambda} (D-2) \int d^D z
 \left[\nabla_{i}\eta\nabla_{j}({N^2\nabla^{j}\zeta}) -\nabla_{i}\zeta \nabla_{j}({N^2\nabla^{j}\eta})\right] {\cal V}'(R)  N^2 \widetilde{C}^i    \no \\
 &&~~~~~~~~~~~~~~~~~~~~~~~~~+\mbox{(higher-derivative contributions)}, \no \\
&&\{ \left<\eta \widetilde{\Omega}\right>,\left<\zeta {\cal H}\right> \}
 \approx-\int d^D z ~(\nabla_{i}\eta) \zeta N ~{\widetilde{\Theta}}^i ,  \no \\
&& \{ \left<\eta \widetilde{\Omega}\right>,\left<\zeta^k {\cal H}_k\right> \}
 =-\int d^D z ~\zeta^k \left[ (\nabla_{k}\eta) ~\widetilde{\Om}-(\nabla_k N^2) ~\widetilde{ C}^i \nabla_i \eta  \right], \no \\
&& \{ \left<\eta \widetilde{\Xi}\right>,\left<\zeta {\cal H}\right> \}
 \approx-\int d^{D} z~ (\nabla_{i}\eta) \zeta N ~ \widetilde{\Sigma}^i +\mbox{(higher-derivative contributions)},  \no \\
&& \{ \left<\eta \widetilde{\Xi}\right>,\left<\zeta^k {\cal H}_k\right> \}
 =-\int d^D z ~ \zeta^k \left[ (\nabla_{k}\eta)   \widetilde{\Xi}-(\nabla_k N^2)~\widetilde{\Theta}^i \nabla_i \eta
 -  \nabla_k N \left< N^2 \left( \f{\de \widetilde{\Theta}^i }{\de N(z)} \right) \nabla_i \eta \right> \right],\no \\
 &&~~~~~~~~~~~~~~~~~~~~~~~~~+\mbox{(higher-derivative contributions)}
 \label{const_Poisson_2_smeared}
\end{\eq}
where,
\begin{\eq}
&&\widetilde{\Sigma}^i =  \left(\frac{\kappa^2}{2}\right) 2 (D-1)  N \left[ \frac{(2\lambda+1) }{(\lambda D -1)} ~\pi~\widetilde{\Theta}^{i} - 2\pi^{i}{}_{k}~\widetilde{\Theta}^{k}\right]  \no \\
&&~~~~~~~~+\left(\frac{\kappa^2}{2}\right)^2 2\xi^2 (D-1) \widehat{\lambda} \left[ \nabla^{i} \mathcal{U}+\frac{ 2 \mathcal{V}^{i}}{(\lambda D-1)} \right]  +\left(\frac{\kappa^2}{2}\right)^{3} \frac{8\xi \widehat{\lambda}}{\sqrt{g}} \mathcal{W}^{i} , \\
&&\mathcal{U} \equiv \left( 2N\nabla_{j}\nabla_{k}N +\nabla_{k}N\nabla_{j}N \right) \left[2 \pi^{jk}-\frac{(2\lambda -1)}{(\lambda D-1)}\pi \, g^{jk} \right]
+ 2 N^{2}\left[\frac{\lambda}{(\lambda D-1)}\pi R -\pi^{jk}R_{jk} \right],\nonumber \\
&&\mathcal{V}^{i} \equiv \pi \left(-N^{2} \nabla^{i}R -\nabla^{i}N \nabla_{j}\nabla^{j}N+N \nabla^{i}\nabla^{j}\nabla_{j}N \right) \no \\
&&~+\nabla^{i}\pi \left[ \frac{ \Lambda (D+2)}{2\xi (D-1)}N^{2} -\frac{(2\lambda -D-1)}{2(D-1)}\left(N^{2}R-N \nabla_{j}\nabla^{j}N \right) \right] \no \\
&&~+ \frac{(\lambda D-1)}{(D-1)}\nabla_{j}\pi \left(N^{2}R^{ij}-N \nabla^{j}\nabla^{i}N \right)
-(\lambda -1)\nabla^{i}(N^{2}\nabla_{j}\nabla^{j}\pi)-\frac{(6\lambda -5)}{2}\nabla^{i}(\nabla^{j}\pi N \nabla_{j}N),\nonumber\\
&&\mathcal{W}^{i} \equiv- \frac{ (2 \lambda+1 ) ( D-2)}{2 ( \lambda D -1)^2}\pi^{3} N\nabla^{i}N- \frac{( 2 \lambda+1 ) (D (2 \lambda +1) -3 \la -2 )}{2 ( \lambda D-1)^2} N^{2}\pi^2\nabla^{i}\pi \no \\
&&~+\frac{(D-2)}{(\lambda D -1)}\pi^{ij}\pi^{2} N\nabla_{j}N-(2D -3)N^{2}\pi^{ik}\pi_{jk} \nabla^{j}\pi+ \frac{(  4D (2 \lambda +1) -12 \la -7)}{2 (\lambda D -1)}N^{2}\pi^{ij}\pi\nabla_{j}\pi. \no \\
\end{\eq}

The Poisson algebra for the local constraints are given by
\begin{\eq}
 &&\{ \widetilde{\Omega}(x),\widetilde{\Omega}(y) \}
 =\left(\frac{\kappa^2}{2}\right)  \widehat{\lambda}(D-2) \nabla_{i}^x
 \left[ {\cal V}'(R) N^2 \widetilde{C}^i (x) \nabla_{j}^y \left(N^2\nabla^{j}_y \de^D (x-y)\right) \right]
 \no \\
 &&~~~~~~~~~~~~~~~~~~~~~~~~~+\mbox{(higher-derivative contributions)} ,  \no \\
&&\{\widetilde{\Omega}(x), {\cal H} (y)\}
 \approx N \widetilde{\Theta}^i (y)\nabla_{i}^x \de^D (x-y),  \no \\
&& \{ \widetilde{\Omega}(x),{\cal H}_k (y) \}
 =-\widetilde{\Om}(y) \nabla_{k}^y \de^D (x-y) - \widetilde{ C}^i  \nabla_i^x
 \de^D (x-y) \nabla_k N^2(y) , \no \\
&& \{ \widetilde{\Xi}(x),{\cal H} (y)\}
 \approx  N \widetilde{\Sigma}^i (y)~ \nabla_{i}^x \de^D (x-y)+\mbox{(higher-derivative contributions)} ,  \no \\
&& \{\widetilde{\Xi}(x),{\cal H}_k (y) \}
 =-\widetilde{\Xi}(y) \nabla_{k}^y \de^D (x-y) - \widetilde{\Theta}^i (y) \nabla_i^x
 \de^D (x-y) \nabla_k N^2(y) \no \\
 &&~~~~~~~~~~~~-  \nabla_i^x  \left[ N^2(x) \left( \f{\de \widetilde{\Theta}^i (x)}{\de N(y)} \right) \right] \nabla_k N(y)+\mbox{(higher-derivative contributions)}  .
 \label{const_Poisson_2_local}
\end{\eq}
Here, we have used
\begin{\eq}
\{ \widetilde{\Omega}(x),\pi_N (y) \}&=& \nabla_{i}^x \left[2 N \widetilde{C}^i (x) \de^D (x-y) \right] ,  \\
 \{ \widetilde{\Xi}(x),\pi_N (y) \}&=& \nabla_{i}^x \left[ 2 N \widetilde{\Theta}^i (x)   \de^D (x-y)
 +  N^2 (x) \left(\f{\de {\Theta}^i (x)}{\de N(y)} \right) \right],
\end{\eq}
where
\begin{\eq}
 &&\left(\f{\de \widetilde{\Theta}^i (x)}{\de N(y)} \right)=-
 \left(\frac{\kappa^{2}}{2}\right)^2 \frac{2 \widehat{\lambda}}{(\lambda D-1) \sqrt{g}} {\cal V}'
 \left\{ \left[ (2\lambda  +1 ) g^{ij}\pi-2(\lambda D-1)\pi^{ij} \right] \nabla_{j}\pi (x)  \de^D (x-y) \right. \no \\
 &&~~\left.+\pi^2 (x) \nabla^i_x \de^D (x-y) \right\}
 -\left(\frac{\kappa^{2}}{2}\right) 2  \widehat{\lambda} (D-1)\sqrt{g} {\cal V}'\nabla^{i}_x
 \left[  \left(\xi R+ \f{\La D}{D-1}  + \frac{D-n}{D-1} \alpha R^{n} \right) \de^D (x-y) \right. \no \\
 &&\left.+\nabla^2 \left(\mathcal{V}^{\prime} \de^D (x-y) \right) \right] - \left(\f{\ka^2}{2} \right) \left[ 2 \widehat{\la} \nabla_i \pi \{{\cal V}'(R)(x), {\cal H} (y)\}+
\{2 (\pi^{ij}-\widehat{\la} g^{ij} \pi )  \nabla_j {\cal V}'(R)(x), {\cal H} (y) \}\right]. \no \\
\end{\eq}

From (\ref{const_Poisson_2_smeared}) or (\ref{const_Poisson_2_local}), one can now compute
\begin{\eq}
\{\widetilde{\Om} , H_C \} &=&\{\widetilde{\Om} , \left< N {\cal H}+N_i {\cal H}^i \right> \} \no \\
 &\approx &\nabla_i \left[ N^2 \widetilde{\Theta}^i  -N_k (\nabla_k N^2) \widetilde{C}^i \right]
 +\nabla_k (N^k \widetilde{\Om}),
 \label{Phi_constraint_Append}
 \\
\{\widetilde{\Xi}(x), H_C \} &=&\{\widetilde{\Phi} (x), \left< N {\cal H}+N_i {\cal H}^i \right> \} \no \\
 &\approx& \nabla_i \left[ N^2 ~\widetilde{\Sigma}^i -N_k (\nabla_k N^2) \widetilde{\Theta}^i  \right]
 +\nabla_k (N^k \widetilde{\Xi}) -\nabla_i^x \left[  N^2(x)  \left<  \left( \f{\de \widetilde{\Theta}^i (x)}{\de N} \right) N^k \nabla_k N \right> \right] \no \\
 &&+\mbox{(higher-derivative contributions)} ,
\label{Phi_time_evol_eq_Append}
\end{\eq}
which reduce to (\ref{Phi_constraint}) and (\ref{Phi_time_evol_final}), respectively, for the case $\widetilde{C}^i = 0$.

\end{section}

\newcommand{\J}[4]{#1 {\bf #2} #3 (#4)}
\newcommand{\andJ}[3]{{\bf #1} (#2) #3}
\newcommand{\AP}{Ann. Phys. (N.Y.)}
\newcommand{\MPL}{Mod. Phys. Lett.}
\newcommand{\NP}{Nucl. Phys.}
\newcommand{\PL}{Phys. Lett.}
\newcommand{\PR}{Phys. Rev. D}
\newcommand{\PRL}{Phys. Rev. Lett.}
\newcommand{\PTP}{Prog. Theor. Phys.}
\newcommand{\hep}[1]{ hep-th/{#1}}
\newcommand{\hepp}[1]{ hep-ph/{#1}}
\newcommand{\hepg}[1]{ gr-qc/{#1}}
\newcommand{\bi}{ \bibitem}

\end{document}